\documentclass[preprint,pof,amsmath,amssymb,aps]{revtex4-1}
\usepackage[abs]{overpic}
\usepackage{graphicx}% Include figure files
\usepackage{dcolumn}% Align table columns on decimal point
\usepackage{bm}% bold math
\usepackage{placeins}
\usepackage{subfig}
\usepackage{color}
\usepackage[normalem]{ulem}
\usepackage{amsmath}
\usepackage{graphicx}
\usepackage{subfig}
\usepackage{amssymb,amsbsy,bm}
\usepackage{mathrsfs,amsmath}
\begin{document}

%\preprint{APS/123-QED} 

\title{Spectral Model of Non-Stationary, Inhomogeneous Turbulence}% Force line breaks with \\

\author{Andrew D. Bragg}
\email{adbragg265@gmail.com}
\author{Susan Kurien}
\affiliation{Applied Mathematics \& Plasma Physics Group, Los Alamos National Laboratory, Los Alamos, NM 87545, USA.}
\author{Timothy T. Clark}
\affiliation{Department of Mechanical Engineering, University of New Mexico, Albuquerque NM USA.}

\date{\today}% It is always \today, today,
       % but any date may be explicitly specified

\begin{abstract}

We compare results from a spectral model for non-stationary, inhomogeneous turbulence (Besnard \emph{et al.}, Theor. Comp. Fluid. Dyn., vol. 8, pp 1-35, 1996) with Direct Numerical Simulation (DNS) data of a shear-free mixing layer (SFML) (Tordella \emph{et al.}, Phys. Rev. E, vol. 77, 016309, 2008). The SFML is used as a test case in which the efficacy of the model closure for the physical-space transport of the fluid velocity field can be tested in a flow with inhomogeneity, without the additional complexity of mean-flow coupling. The model is able to capture certain features of the SFML quite well for intermediate to long-times, including the evolution of the mixing-layer width and turbulent kinetic energy. At short-times, and for more sensitive statistics such as the generation of the velocity field anisotropy, the model is less accurate. We present arguments, supported by the DNS data, that a significant cause of the discrepancies is the local approximation to the intrinsically non-local pressure-transport in physical-space that was made in the model, the effects of which would be particularly strong at short-times when the inhomogeneity of the SFML is strongest.
\end{abstract}

\maketitle

%\tableofcontents

%\section{First section}
%
%Article content.
%
%\subsection{A Subsection}

\section{Introduction}

Developing theoretical models that can predict the statistical features of inhomogeneous, turbulent flows presents a major challenge, and there are several possible approaches and perspectives to consider. One important choice is to determine the space in which one wishes to describe the turbulence statistics, e.g. a `one-point' space (describing the turbulent statistics at one point in space $\bm{x}$) or a `multi-point' space (describing the turbulent statistics at multiple, distinct points in space $\bm{x}_1, \bm{x}_2...\bm{x}_n$). This choice is largely determined by the desired balance between the solvability of the model and the level of faithfulness one wishes to retain to the true turbulent dynamics. The majority of modeling efforts have tended to focus on one-point models, often in the context of the so-called `Two Equation Models' (see Pope \cite{pope}). However, from a fundamental perspective, the models should at least describe the turbulence at the two-point level. This is both because the incompressible Navier-Stokes Equation (NSE) is itself \emph{essentially} a two-point system, since the pressure field is non-local in physical-space, and also because processes such as the energy transfer among the scales of motion in turbulence cannot, by definition, be described with less than two-points since a single point carries no information on scale.

The next important choice is to decide whether one wishes to construct the model deductively, from the NSE, or to construct it phenomenologically. As for other non-linear, field-theoretic problems in physics, one could seek to construct a theoretical model directly from the governing dynamical equation (in this case the NSE) by using renormalized perturbation theory (RPT). In the context of turbulence, RPT can be viewed as the resummation of certain classes of diagrams that arise from a diagrammatic representation of the primitive perturbation expansion of the NSE \cite{berera13}. Kraichnan's Direct Interaction Approximation \cite{kraichnan59} can be viewed as a RPT, as can the theory of Wyld \cite{wyld61} and that of Martin, Siggia and Rose \cite{martin73}. The use of RPT to construct theories of turbulence has been quite successful for the simplified case of homogeneous, isotropic turbulence, providing quite accurate predictions for some of the statistics of the turbulent velocity field (see McComb \& Quinn \cite{mccomb03}). However, as is well known, many of the RPTs are inconsistent with known behavior of the turbulent velocity field in the inertial range, and in fact the only purely Eulerian theory that is consistent with the known inertial range behavior is the Local Energy Transfer theory of McComb \cite{mccomb74}. Although it is possible to construct RPTs for inhomogeneous turbulence, the resulting theories are very complex, and the equations would be far too difficult to solve even with current computing power. 

An alternative approach to RPT is to construct the turbulence model phenomenologically, in which the non-linear and non-local processes in turbulence are represented by terms in the equations that reproduce the correct, known qualitative behavior. The disadvantage with this approach, aside from being unsatisfying from a fundamental perspective, is that it introduces unknown constants into the model. Sometimes the constants can be specified theoretically by appealing to certain asymptotic constraints on the system, but sometimes they simply have to be obtained by fitting the model to numerical or experimental data. Nevertheless, for complex turbulent flows, a phenomenological approach is often the only feasible option.

In this paper we consider the phenomenological turbulence model proposed by Besnard et al. \cite{bhrz96} (referred to hereafter as the BHRZ model), which is in principle able to describe statistically non-stationary, inhomogeneous and anisotropic turbulent flows. The modeling philosophy behind the BHRZ model is similar in spirit to some of the earlier models such as that of Daly \& Harlow \cite{daly70} and that of Launder, Reece \& Rodi \cite{launder75}. However, a crucial difference is that whereas these models are one-point models, the BHRZ model is a two-point model. Some of the advantages of a two-point model compared with a one-point model is that the former, unlike the latter, can describe the evolution of the distribution of energy among the scales of motion of the turbulence, which is essential for correctly handling non-stationary flows. In addition to this, a two-point model removes the need to specify an equation for the evolution of the turbulent kinetic energy dissipation-rate, the construction of which usually involves the introduction of many approximations that are especially unsuitable for non-stationary, inhomogeneous turbulent flows. 
 
Previous comparisons of the BHRZ model with homogeneous sheared turbulence in Clark \& Zemach \cite{clark95} showed good agreement, and that the model was able to capture certain non-trivial aspects of the flow. Detailed investigations of the validity of the BHRZ model for predicting inhomogeneous turbulent flows have not, however, been undertaken, and this is precisely the purpose of the present paper. It is the way that the BHRZ models the physical-space transport of the turbulent velocity fluctuations that is of primary concern, since the model made several approximations in describing this processes. The closure model employed in BHRZ for the physical-space transport also contains an unknown constant; one of the aims of the present work is to obtain an estimate for this constant.

Rather than considering a more complex flow such as a turbulent boundary layer, we instead compare the BHRZ model with DNS data of a Shear-Free Mixing Layer (SFML), explained in detail in \S\ref{SFML_setup}. One reason for this choice of flow configuration is that it allows us to focus on the BHRZ closure model used for the physical-space transport of the turbulent velocity fluctuations, without the obscuring effect of mean-flow coupling. 

The outline of the paper is as follows: In \S\ref{SFML_setup} we explain the SFML, the physical-mechanisms that govern its evolution, and the particular statistical features of the flow. Next, in \S\ref{BHRZmodel} we introduce the BHRZ model and discuss in detail the various approximations made in its derivation. We also discuss the initial and boundary conditions used in the model for the SFML. In \S\ref{ResDis} we compare the predictions from the BHRZ model with Direct Numerical Simulation (DNS) data of a SFML. Finally, in \S\ref{Conc} we draw conclusions to the work and identify features of the BHRZ model than need to be improved in future work in order for it to more accurately predict strongly non-stationary, inhomogeneous and anisotropic turbulent flows.

\section{Shear-Free Mixing Layer}\label{SFML_setup}

In this section we first describe the SFML simulated in Tordella, Iovieno \& Bailey \cite{tordella08}, against which we shall compare the BHRZ model predictions in \S\ref{ResDis}, and then consider the physical mechanisms governing its evolution and the statistical characteristics of the flow.

\begin{figure}[ht]
\centering
{\begin{overpic}
[trim = 0mm 65mm 0mm 70mm,scale=0.4,clip,tics=20]{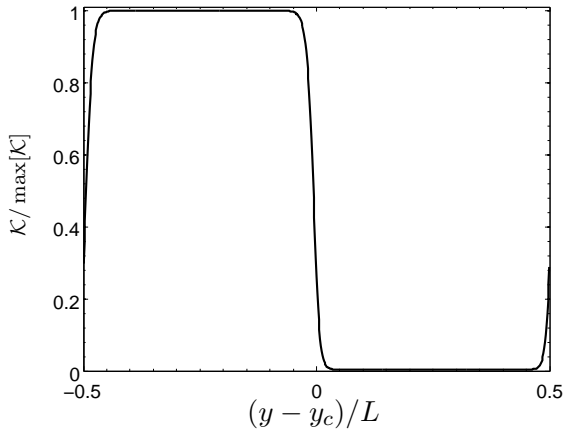}
\put(100,0){$(y-y_c)/L$}
\put(10,70){\rotatebox{90}{\scriptsize{$\mathcal{K}/\max[\mathcal{K}]$}}}
\end{overpic}}
\caption{Illustration of the turbulent kinetic energy, $\mathcal{K}(y,t)$, in a SFML, where ${y\in[0,L]}$, $L$ is the periodic lengthscale of $\mathcal{K}(y,t)$ and $y_c\equiv L/2$.}
\label{SFML_diag}
\end{figure}

Figure~\ref{SFML_diag} illustrates the turbulent kinetic energy (TKE), $\mathcal{K}(y,t)$, in the SFML. The function satisfies $\mathcal{K}(y,t)=\mathcal{K}(y+L,t)$ where ${y\in[0,L]}$ and $L$ is the periodic lengthscale. At ${t=0}$ the ratio of the maximum to the minimum TKE in the SFML is given by $\max[\mathcal{K}(y,0)]/\min[\mathcal{K}(y,0)]$, and by varying this ratio, the strength of the initial inhomogeneity in the SFML can be controlled, and its effect upon the resulting flow can be examined. Note that in Tordella \emph{et al.} \cite{tordella08}, only the TKE is inhomogeneous in the initial flow field; the integral lengthscale is homogeneous.
\FloatBarrier
In Tordella \emph{et al.} \cite{tordella08}, the initial velocity field for the SFML is constructed as follows: Let $\bm{u}^{[1]}(\bm{x},t)$ be a Navier-Stokes turbulent velocity field that is statistically stationary, homogeneous and isotropic. Now define a second velocity field $\bm{u}^{[2]}(\bm{x},t)\equiv\gamma\bm{u}^{[1]}(\bm{x},t)$, where $\gamma$ is a constant $\leq1$. The initial velocity field for the SFML is then constructed as
\begin{align}
\bm{u}(\bm{x},0)\equiv P(y)\bm{u}^{[1]}(\bm{x},0)+\Big(1-P(y)\Big)\bm{u}^{[2]}(\bm{x},0),
\end{align}
where $y\equiv x_2$ is the inhomogeneous, mixing direction. The function $P(y)\in[0,1]$, such that in the `low energy region', $\bm{u}^{[2]}(\bm{x},0)$ describes the initial velocity field, and in the `high energy region', $\bm{u}^{[1]}(\bm{x},0)$ describes the initial velocity field, and satisfies $P(y)=P(y+L)$. The particular form of $P(y)$ used in Tordella \emph{et al.} \cite{tordella08} is
\begin{align}
P(y)=\frac{1}{2}\Big(1+\mathrm{tanh}\Big[a y/L\Big]\mathrm{tanh}\Big[a (y-L/2)/L\Big]\mathrm{tanh}\Big[a (y-L)/L\Big]\Big),
\end{align}
with $a=20\pi$ and $L=4\pi$. For each value of $\gamma$ chosen, $\bm{u}^{[1]}(\bm{x},0)$ is fixed. This implies that the total energy in the mixing layer varies with $\gamma$. However, importantly, it also means that the high energy region of the mixing layer has an initial TKE that is independent of $\gamma$ and so its rate of energy dissipation does not vary with $\gamma$. This ensures that as $\gamma$ is varied, the variation in the rate at which energy initially leaves the high energy region of the mixing layer is governed solely by the variation of the strength of the physical-space energy transport with $\gamma$.

The Reynolds stress tensor for $\bm{u}(\bm{x},0)$ is
\begin{align}
\begin{split}
\Big\langle\bm{u}(\bm{x},0)\bm{u}(\bm{x},0)\Big\rangle\equiv &P^2\Big\langle\bm{u}^{[1]}(\bm{x},0)\bm{u}^{[1]}(\bm{x},0)\Big\rangle+P\Big(1-P\Big)\Big\langle\bm{u}^{[1]}(\bm{x},0)\bm{u}^{[2]}(\bm{x},0)\Big\rangle\\
&+P\Big(1-P\Big)\Big\langle\bm{u}^{[2]}(\bm{x},0)\bm{u}^{[1]}(\bm{x},0)\Big\rangle+\Big(1-P\Big)^2\Big\langle\bm{u}^{[2]}(\bm{x},0)\bm{u}^{[2]}(\bm{x},0)\Big\rangle,
\end{split}
\end{align}
where $\langle\cdot\rangle$ denotes an ensemble average, and invoking the statistical properties and definitions of $\bm{u}^{[1]}$ and $\bm{u}^{[2]}$ we have
\begin{align}
\Big\langle\bm{u}(y,0)\bm{u}(y,0)\Big\rangle&=(2/3)\mathcal{K}^{[1]}\Big(P^2 +\gamma[1-P]\Big(2 P+\gamma[1-P]\Big)\Big)\mathbf{I},\label{uuinit}\\
\mathcal{K}(y,0)&=\mathcal{K}^{[1]}\Big(P^2 +\gamma[1-P]\Big(2 P+\gamma[1-P]\Big)\Big),\label{Kinit}
\end{align}
where $\mathcal{K}^{[1]}(y,0)\equiv (1/2)\mathrm{tr}[\langle\bm{u}^{[1]}(y,0)\bm{u}^{[1]}(y,0)\rangle]$ and $\mathbf{I}$ is the identity tensor. We also note that $\gamma=\sqrt{\min[\mathcal{K}(y,0)]/\max[\mathcal{K}(y,0)]}$ which will be used as a control parameter in \S\ref{ResDis} to vary the strength of the initial inhomogeneity of the SFML. 

Having defined the construction of $\mathcal{K}(y,0)$, we now consider how $\mathcal{K}(y,t)$ evolves as the flow mixes. With the initial condition (\ref{uuinit}) the flow is inhomogeneous in the $y$ direction, and consequently the transport mechanisms in the flow proceed to mix the TKE. The flow remains homogeneous in the $(x_1,x_2)$ directions, such that the flow exhibits cylindrical symmetry about the $y\equiv x_2$ axis. In Tordella \emph{et al.} \cite{tordella08} this fact is exploited for the construction of the statistics of the flow field which are constructed as averages over the $(x_1,x_2)$-plane, for a given $y,t$. The transport equation governing $\mathcal{K}(y,t)$ in the SFML is given by 
\begin{align}
\partial_t\mathcal{K}(y,t)=-\underbrace{(1/2)\partial_y\mathrm{tr}[\langle\bm{uu}u_2\rangle]}_{\text{Self-transport}}-\underbrace{(1/2)\partial_y\langle u_2 p\rangle}_{\text{Pressure-transport}}+\underbrace{(\nu/2)\partial_y^2\mathcal{K}}_{\text{Molecular transport}}-\underbrace{\nu\langle\bm{\partial_x u:\partial_x u}\rangle,}_{\text{Dissipation}}\label{TKEeqn}
\end{align}
where $\mathcal{K}(y,t)\equiv(1/2)\mathrm{tr}[\langle\bm{u}(y,t)\bm{u}(y,t)\rangle]$, $\bm{u}(\bm{x},t)$ is the fluid velocity field, $p(\bm{x},t)$ is the pressure field,  and $\nu$ is the kinematic viscosity of the fluid. Note that in the SFML that we are considering, $\langle\bm{u}(\bm{x},t)\rangle=\bm{0}\,\forall t$ and therefore $\bm{u}(\bm{x},t)$ represents the fluctuating velocity field. 

When $\bm{u}(\bm{x},t)$ is turbulent, the contribution from the self-transport and pressure-transport terms will typically dominate the overall transport of $\mathcal{K}$ in physical-space, with molecular diffusion transport playing a small role. For the SFML that we are considering, the initial velocity field satisfies $\langle\bm{u}(y,0) \bm{u}(y,0) \rangle\propto\mathbf{I}$. With this initial condition, it is straightforward to show that the self-transport and pressure-transport terms in the equation governing the evolution of $\langle\bm{u}(y,t)\bm{u}(y,t)\rangle$ are the only terms that are able to generate anisotropy in the flow field. It is this, combined with the dominance of the self-transport and pressure-transport terms in the mixing process, that makes the SFML a good test case for the BHRZ model since it provides a flow configuration for which the BHRZ closure modeling for the physical-space transport of $\bm{u}(\bm{x},t)$ can be scrutinized.

One of the significant complexities of inhomogeneous turbulent flows is that the physical-space transport is non-local because of the contribution from the pressure field. Although the pressure field is dynamically non-local in any turbulent flow, it only makes an explicit, finite statistical contribution to the evolution of $\mathcal{K}(y,t)$ when the flow is inhomogeneous. This may be seen by noting that for an incompressible turbulent flow, $(\bm{\partial_x\cdot\partial_x}) p(\bm{x},t)=-\bm{\partial_x\partial_x:}[\bm{u}(\bm{x},t)\bm{u}(\bm{x},t)]$, and by expressing the solution to this Poisson equation using a Green function, $G$, we may write
\begin{align}
\bm{\partial_x}\langle\bm{u}(\bm{x},t)p(\bm{x},t)\rangle=-\bm{\partial_x}\int\limits_{\mathbb{R}^3}G(\bm{x},\bm{x}')\bm{\partial_{x'}\partial_{x'}:}\Big\langle\bm{u}(\bm{x},t)\bm{u}(\bm{x}',t)\bm{u}(\bm{x}',t)\Big\rangle\, d\bm{x}'.
\end{align}
The complexity that this non-locality introduces is problematic both computationally, since it renders the evolution equation for $\mathcal{K}(y,t)$ of integro-differential form, and also theoretically, because it requires knowledge of the two-point statistic $\langle u_2(y,t)u_2(y',t)u_2(y',t)\rangle$. In contrast to turbulent flows with a mean velocity field, in Tordella \emph{et al.} \cite{tordella08} it was observed that for the SFML, the self-transport does not dominate the pressure-transport. Consequently, in constructing models for the SFML, errors incurred by making approximations to the pressure-transport (such as a local approximation) could have a significant effect on the overall predictions of the model for the SFML. We will return to this topic throughout the paper, in considering how the BHRZ model closes the pressure-transport contribution and also in considering the effects of the closure model on the ability of BHRZ to accurately describe inhomogeneous flows.

We now turn to consider some of the statistical features of the SFML during its evolution. The initial inhomogeneity of $\mathcal{K}$ leads to a mixing of the TKE of the flow through the action of the transport mechanisms described by (\ref{TKEeqn}), and these transport mechanisms drive the velocity field towards the asymptotically homogeneous state $\lim_{t\to\infty}(\max[\mathcal{K}(y,t)]/\min[\mathcal{K}(y,t)])\to1$. The results in Tordella \emph{et al.} \cite{tordella08} showed that after a time period of $t/\tau\approx 3$, where $\tau$ is the integral timescale of the initial field $\bm{u}^{[1]}(\bm{x},t)$, the SFML enters a self-similar regime, during which the statistics of $\bm{u}(\bm{x},t)$ become approximately constant when normalized in appropriate ways. The self-similar regime corresponds to the regime where the competition between the physical mechanisms governing the evolution have come to a sort of `equilibrium'. For example, in the self-similar regime the flow field anisotropy is approximately constant, indicating that the transport mechanisms generating the anisotropy have reached an equilibrium state where they balance out the return to isotropy processes acting in the field. It is in fact only the transport mechanisms that can break the isotropic symmetry of the initial field $\bm{u}(\bm{x},0)$ as the mixing occurs, since there is no mean-shear in the SFML. The transport terms break the isotropic symmetry of the velocity field because different components of the velocity field are transported at different rates. For example, the self-transport is $-\partial_y\langle\bm{u}(y,t)\bm{u}(y,t)u_2(y,t)\rangle$ and its contribution to $\partial_t\langle u_1(y,t)u_1(y,t)\rangle$ and $\partial_t\langle u_2(y,t)u_2(y,t)\rangle$ will differ since $\langle u_1(y,t)u_1(y,t)u_2(y,t)\rangle\neq\langle u_2(y,t)u_2(y,t)u_2(y,t)\rangle$ (since $u_2$ is more correlated with itself than with $u_1$).

In an early experimental study on the SFML, Gilbert \cite{gilbert80} found that the Probability Density Function (PDF) of $\bm{u}(\bm{x},t)$ remained essentially Gaussian throughout the evolution of the mixing process. In subsequent experiments on the SFML performed by Veeravalli \& Warhaft \cite{veeravalli89}, a very different behavior was observed, with $\bm{u}(\bm{x},t)$ exhibiting strong departures from Gaussianity. Veeravalli \& Warhaft attributed this to the fact that in Gilbert's experiments, $\max[\mathcal{K}(y,0)]/\min[\mathcal{K}(y,0)]$ was too low ($\approx 1.48$) for the transport mechanisms to substantially drive the PDF of $\bm{u}(\bm{x},t)$ away from its initial Gaussian form. In both Gilbert's and Veeravalli \& Warhaft's experiments on the SFML, there was an inhomogeneity in both the initial TKE and the initial integral lengthscale across the mixing layer. 

The purpose of the study in Tordella \emph{et al.} \cite{tordella08} was to demonstrate that an inhomogeneity in the initial TKE alone is a sufficient condition for the  PDF of $\bm{u}(\bm{x},t)$ to deviate from its Gaussian form at $t=0$. They observed strong departures from Gaussianity, and that the field became more non-Gaussian with increasing $\max[\mathcal{K}(y,0)]/\min[\mathcal{K}(y,0)]$, until the field reached a saturation point at $\max[\mathcal{K}(y,0)]/\min[\mathcal{K}(y,0)]=\mathcal{O}(100)$ above which the non-Gaussianity did not continue to grow. The maximum skewness of $\bm{u}(\bm{x},t)$ in the mixing layer reached an approximately constant value of $1$ at long times for $\max[\mathcal{K}(y,0)]/\min[\mathcal{K}(y,0)]=12$, and reached an approximately constant value of $2.25$ at long times for $\max[\mathcal{K}(y,0)]/\min[\mathcal{K}(y,0)]=300$. The maximum kurtosis of $\bm{u}(\bm{x},t)$ in the mixing layer reached an approximately constant value of $4$ at long times for $\max[\mathcal{K}(y,0)]/\min[\mathcal{K}(y,0)]=12$, and reached an approximately constant value of $11$ at long times for $\max[\mathcal{K}(y,0)]/\min[\mathcal{K}(y,0)]=300$. Besides the importance of these results for fundamental reasons, these results also have important implications for the construction of models for the SFML. In particular, the strong non-Gaussianity of the velocity field in the SFML may be problematic since the vast majority of models (including the BHRZ model, as discussed in \S\ref{BHRZmodel}) for inhomogeneous turbulent flows make closure approximations that are only strictly justifiable in the limit where the PDF of $\bm{u}(\bm{x},t)$ is only weakly perturbed from a Gaussian.

\section{BHRZ model for the SFML}\label{BHRZmodel}

We shall now introduce the BHRZ model and consider in detail the various assumptions that have been made in its construction. Such a detailed presentation of the model will prove to be useful when we come to compare its predictions with the DNS data for the SFML in \S\ref{ResDis}. We then consider the specification of certain constants appearing in the model, and then present the initial and boundary conditions used in the BHRZ model for the SFML. It is important to note that what we are here calling BHRZ actually differs in some respects to the original BHRZ model presented in Besnard \emph{et al.} \cite{bhrz96}, in particular, the specification of the timescale of the turbulent scales of motion and the symmetrization of the physical-space transport term (both of these differences are discussed below). Nevertheless, we retain the name BHRZ since that is the origin of the essential modeling framework that we are considered here.

\subsection{Formulation of the model}

The BHRZ model begins by considering $\bm{\mathcal{R}}_2(\bm{x}_1,\bm{x}_2,t)\equiv\langle\bm{u}(\bm{x}_1,t)\bm{u}(\bm{x}_2,t)\rangle$, the two-point, one-time correlation tensor of the fluid velocity field $\bm{u}(\bm{x},t)$, for which an evolution equation can be obtained from the incompressible NSE. Applying the transformation $[\bm{x}_1,\bm{x}_2]\to[\bm{x}+\bm{r}/2,\bm{x}-\bm{r}/2]$ to the evolution equation for $\bm{\mathcal{R}}_2(\bm{x}_1,\bm{x}_2,t)$, and then applying a Fourier transform conjugate to $\bm{r}$, we obtain (for $\langle\bm{u}(\bm{x},t)\rangle=\bm{0}$)
\begin{align}
\partial_t \bm{\mathcal{R}}_2(\bm{x},\bm{k},t)=	{\mathcal{V}}\bm{\mathcal{R}}_2+\bm{\mathcal{N}}+\bm{\mathcal{N}}^\dag,\label{R2eqn}
\end{align}
where `$\dag$' denotes the Hermitian,
\begin{align}
{\mathcal{V}}(\bm{x},\bm{k})&\equiv-2\nu k^2+(\nu/2)\partial_{x}^2,\\
\begin{split}
\mathcal{N}_{ij}(\bm{x},\bm{k},t)&\equiv-\nabla_n \mathcal{R}_{3,inj}(\bm{x},\bm{k},t)-\nabla_i\int\limits_{\mathbb{R}^3} G(\bm{x},\bm{x}',\bm{k}) \nabla'_m\nabla'_n \mathcal{R}_{3,mnj}(\bm{x}',\bm{k},t)\,d\bm{x}',\label{Nft}\\
 \mathcal{R}_{2,ij}(\bm{x},\bm{k},t)&\equiv\int\limits_{\mathbb{R}^3} e^{-i\bm{k\cdot r}}\Big\langle u_i(\bm{x}+\bm{r}/2,t)u_j(\bm{x}-\bm{r}/2,t)\Big\rangle\,d\bm{r},\\
 \mathcal{R}_{3,inj}(\bm{x},\bm{k},t)&\equiv\int\limits_{\mathbb{R}^3} e^{-i\bm{k\cdot r}}\Big\langle u_i(\bm{x}+\bm{r}/2,t)u_n(\bm{x}+\bm{r}/2,t) u_j(\bm{x}-\bm{r}/2,t)\Big\rangle\,d\bm{r},\\
{\nabla_m}&\equiv\frac{1}{2}\frac{\partial}{\partial{x}_m}+i{k}_m,
\end{split}
\end{align}
and the Green function, $G(\bm{x},\bm{x}',\bm{k})$, solves 
\begin{align}
-\nabla^2 G(\bm{x},\bm{x}',\bm{k})=\delta(\bm{x}-\bm{x}'),\quad\text{for}\,\,\bm{x}\in\mathbb{R}^3.	
\end{align}
The term ${\mathcal{V}}\bm{\mathcal{R}}_2$ in (\ref{R2eqn}) describes the contribution to the evolution of $\bm{\mathcal{R}}_2$ from molecular dissipation and diffusion, and is in closed form. The term $\bm{\mathcal{N}}$ is unclosed and contains contributions arising from two distinct terms in the NSE; the first coming from the nonlinear advection term, and the second from the pressure gradient term. The second contribution is non-local in physical-space ($\bm{x}$-space) and makes the evolution equation for $\bm{\mathcal{R}}_2$ an integro-differential equation. The two contributions in $\bm{\mathcal{N}}$ collectively describe three physically distinct processes affecting the evolution of $\bm{\mathcal{R}}_2$; a redistribution effect, energy transport in $\bm{k}$-space and energy transport in $\bm{x}$-space. The BHRZ model closes $\bm{\mathcal{N}}$ phenomenologically, utilizing closures that qualitatively reproduce each of these physical processes. 

The first step made in BHRZ is to integrate the evolution equation for $\bm{\mathcal{R}}_2(\bm{x},\bm{k},t)$ over spherical shells of constant $k\equiv\|\bm{k}\|$, defining
\begin{align}
\bm{E}(\bm{x},{k},t)\equiv\int\bm{ \mathcal{R}}_2(\bm{x},\bm{k},t)\frac{k^2}{(2\pi)^3}\,d\Omega_k,
\end{align}
such that 
\begin{align}
\mathcal{K}(\bm{x},t)\equiv\int_0^{\infty}\mathrm{tr}[\bm{E}(\bm{x},k,t)]\,dk,
\end{align}
is the TKE per unit mass at $(\bm{x},t)$ and
\begin{align}
\bm{R}_{2}(\bm{x},t)\equiv\Big\langle\bm{u}(\bm{x},t)\bm{u}(\bm{x},t)\Big\rangle=2\int_0^{\infty}\bm{E}(\bm{x},k,t)\,dk,	
\end{align}
is the one-point Reynolds stress tensor. Note that here and throughout, $\bm{\mathcal{R}}_N$, is used to denote the $N^{th}$ order velocity correlation tensors in $(\bm{x},\bm{k},t)$-space, whereas $\bm{R}_N$ is used to denote them in $(\bm{x},t)$-space.

As discussed in Besnard \emph{et al.} \cite{bhrz96}, the integration over spherical shells (``angle averaging'') is introduced merely to reduce the dimensionality of the solution space of the model. Such a procedure is without justification in the general case of anisotropic velocity fields, and impacts the ability of the model to represent anisotropic dynamics in $\bm{k}$-space. The precise way in which this angle averaging affects the ability of the model to accurately predict the evolution of $\bm{\mathcal{R}}_2$ is not entirely understood. However, recent work has made a step towards understanding these effects by constructing tensor spherical harmonic expansions for objects such as $\bm{\mathcal{R}}_2$ (in homogeneous flows) that can be used to consider the evolution and contribution of the terms neglected through the spherical average to the evolution of $\bm{\mathcal{R}}_2$ (see Rubinstein, Kurien \& Cambon \cite{rubinstein15}). It should be noted, however, that this spherical averaging only directly suppresses the anisotropy description in $\bm{k}$-space, that is, in scale, and not $\bm{x}$-space, that is, in position. This is important because the transport of $\bm{\mathcal{R}}_2$ that dominates the mixing in the SFML and generates the flow field anisotropy is dominated by the motion of the largest scales of the flow, and this transport occurs in $\bm{x}$-space. Thus, much of the anisotropy of the flow that is important for the mixing process is still captured through the models description of anisotropy in $\bm{x}$-space.

With the spherical averaging operation applied to the evolution equation for $\bm{\mathcal{R}}_2$, phenomenological closure models are then introduced to capture the distinct physical processes described by $\bm{\mathcal{N}}(\bm{x},{k},t)$. In the following, for simplicity, we only discuss the form of the BHRZ closures that are applicable to the SFML, for which the terms in the evolution equation for $\bm{\mathcal{R}}_2$ are described by the reduced variables $(y,k,t)\equiv(x_2,k,t)$. 

The redistribution effect in $\bm{\mathcal{N}}$ does not contribute to the evolution of $\mathcal{K}$, but instead redistributes the energy among the different components of the velocity field. For example, in a one-point equation for $\bm{R}_{2}$, the redistribution effect in a homogeneous turbulent velocity field is described by $\langle p(\bm{x},t)\bm{\nabla u}(\bm{x},t)\rangle$, which does not contribute to $\mathcal{K}(\bm{x},t)$ in an incompressible velocity field since $\mathrm{tr}[\langle p(\bm{x},t)\bm{\nabla u}(\bm{x},t)\rangle]=\bm{0}$. It is common to assume, especially when constructing models, that the nature of the redistribution effect is to drive $\bm{u}(\bm{x},t)$ towards an isotropic state (though this need not be the case), and following this line of thought, BHRZ models the redistribution effect with a linear return-to-isotropy term
\begin{align}
\bm{\mathcal{A}}(y,k,t)\equiv c_M\Phi\Big((E/3)\bm{\mathrm{I}}-\bm{E}\Big),
\end{align}
where $c_M$ is a dimensionless constant (the model constrants introduced in this section are discussed in \S\ref{DMC}), $\Phi(y,k,t)$ is a frequency (defined below) and $E(y,k,t)\equiv\mathrm{tr}[\bm{E}(y,k,t)]$. This is the simplest model for the return-to-isotropy tensor, and ignores contributions involving nonlinear combinations of the anisotropy measure $(E/3)\bm{\mathrm{I}}-\bm{E}$. In the context of one-point turbulence models where the return-to-isotropy term is proportional to $(\mathrm{tr}[\bm{R}_2]/3)\bm{\mathrm{I}}-\bm{R}_2$, such linear return-to-isotropy models are known to be qualitatively incorrect \cite{lumley77}. However, its use in BHRZ allows for a more faithful representation of the return-to-isotropy process since in this case the model is able to capture the fact that different scales of motion in the turbulence return-to-isotropy at different rates, according to the correlation timescale of their motion. Indeed, in Clark \& Zemach \cite{clark95} it was found that the BHRZ model was able to capture the evolution of the flow anisotropy quite well, for the case of a homogeneous shear flow.
 
The transport in ${k}$-space is modeled using an advection-diffusion equation of the form proposed by Leith \cite{leith67}
\begin{align}
\bm{\mathcal{T}}^k(y,k,t)\equiv -c_1\partial_k\Big(k\Phi\bm{E}\Big)+c_2\partial_k\Big(k^2\Phi\partial_k\bm{E}\Big),
\end{align}
where $c_1$ and $c_2$ are dimensionless constants. The use of such a model approximates the ${k}$-space transport as being local in ${k}$-space, and indeed in Clark, Rubinstein \& Weinstock \cite{clark09} it was shown that the structure of the Leith diffusion model can be understood to arise from restricting the triad interactions in RPTs (such as the DIA) to local interactions. Such a local truncation is in principle inconsistent with the NSE that describes non-local interactions among the scales of the turbulent velocity field. However, there are rigorous theoretical and numerical results that support the idea that the energy transport among scales in the inertial range of homogeneous, isotropic turbulence is dominated by local scale interactions \cite{aluie09}. This at least suggests that in the limit of weak inhomogeneity, the BHRZ model for the ${k}$-space energy transport is reasonable. Comparisons of results from the BHRZ model to results from the non-local EDQNM model in Besnard \emph{et al.} \cite{bhrz96} showed close agreement, also suggesting that the non-local interactions play a small role. 

The transport in $\bm{x}$-space is modeled in BHRZ in essentially two steps: First, a local approximation is made to the non-local (in $\bm{x}$-space) pressure-transport in $\bm{\mathcal{N}}$, which amounts to setting $G(\bm{x},\bm{x}',\bm{k})\bm{\nabla}'\bm{\nabla}'\approx \delta(\bm{x}-\bm{x}')k^{-2}\bm{kk}$ in (\ref{Nft}). The resulting $\bm{x}$-space transport terms involve $\bm{\mathcal{R}}_3$, and this is closed by using a gradient-diffusion approximation, wherein $\bm{\mathcal{R}}_3$ is related to the gradients of $\bm{\mathcal{R}}_2$, giving the spherically averaged result (in component form)
\begin{align}
\mathcal{T}^y_{ij}(y,k,t)\equiv c_D\partial_{y}\Big(\mathscr{D}_{22}\partial_{y}{E}_{ij}+\mathscr{D}_{j2}\partial_{y}{E}_{2i}+\mathscr{D}_{i2}\partial_{y}{E}_{j2}\Big),\label{Tx}
\end{align}
where $c_D$ is a dimensionless constant and $\bm{\mathscr{D}}(y,t)$ is a diffusion tensor, defined as
\begin{align}
	\bm{\mathscr{D}}(y,t)\equiv \int\limits_0^\infty \Phi^{-1}(y,q,t)\bm{E}(y,q,t)\,dq.
\end{align}
%. 
The form of the closure in (\ref{Tx}) differs from that proposed in the original BHRZ model: In Besnard \emph{et al.} \cite{bhrz96} they discuss a closure of the form given in (\ref{Tx}) but subsequently choose to adopt a simpler closure $\bm{\mathcal{T}}^y=c_D\partial_y(\nu_T \partial_y\bm{E})$. However, this simplified form does not correctly preserve the component coupling of $\bm{u}(\bm{x},t)$, and for the SFML with initially isotropic $\bm{E}$, the use of the simplified model for $\bm{\mathcal{T}}^y$ in BHRZ would lead to the incorrect prediction that $\bm{E}(y,k,t)\propto\mathbf{I}\,\forall t$.

For strongly inhomogeneous turbulent flows, the local approximation to the pressure-transport in $\bm{x}$-space is likely in significant error, since in such a flow, $\bm{\mathcal{R}}_3$ will vary considerably over the range of the support of the integrand describing the pressure-transport. According to the analysis in Aluie \& Eyink \cite{aluie09b}, the locality of the $\bm{k}$-space energy transport essentially follows from the particular scaling properties of the Fourier transform of the velocity field, i.e. $\bm{u}(\bm{k},t)$, for $\bm{k}$ in the inertial range, and these scalings are to some degree universal. In contrast, the $\bm{x}$-space energy transport depends mainly upon the motion of the largest scales of the turbulence, whose scaling is fundamentally different to that of the velocity differences in the inertial range, and furthermore, their behavior is entirely problem dependent (i.e. definitely not universal). Therefore, whereas locality of the transport in $\bm{k}$-space may hold (even approximately) for $\bm{k}$ in the inertial range of an inhomogeneous turbulent flow in the limit $Re_\lambda\to\infty$, there is no corresponding reason why there should be locality of the pressure-transport in $\bm{x}$-space for an incompressible flow. 

Concerning the gradient-diffusion closure employed for $\bm{\mathcal{R}}_3$, this is only strictly applicable in the limit of weak inhomogeneity, where the  PDF of $\bm{u}(\bm{x},t)$ is only weakly perturbed from a Gaussian (noting that the  PDF of $\bm{u}(\bm{x},t)$ is essentially Gaussian in homogeneous turbulence). 

With these phenomenological closures, the BHRZ model for $\bm{E}(y,{k},t)$ in the SFML is
\begin{align}
\partial_t\bm{E}(y,{k},t)\approx\mathcal{V}\bm{R}_2+\bm{\mathcal{A}}+\bm{\mathcal{T}}^y+\bm{\mathcal{T}}^k.\label{BHRZ_closure_N}		
\end{align}
Each of the closure models $\bm{\mathcal{A}}$, $\bm{\mathcal{T}}^y$ and $\bm{\mathcal{T}}^k$ involve the frequency $\Phi(y,k,t)$, and in Besnard \emph{et al.} \cite{bhrz96} the simple form $\Phi(y,k,t)=\sqrt{k^3 E(y,k,t)}$ is utilized. We instead choose the form for $\Phi(y,k,t)$ suggested in Rubinstein \& Clark \cite{rubinstein04} that captures the viscous effects on the frequency (which we found to be important for describing the low Reynolds number SFML considered in Tordella \emph{et al.} \cite{tordella08}), namely
\begin{align}
\Phi(y,k,t)&=(1/2)\Big[\Big(\nu^2 k^4+4H\Big)^{1/2}-\nu k^2\Big],\\
H(y,k,t)&\equiv c_H\int\limits_0^k q^2 E(y,q,t)\,dq,
\end{align}
where $c_H=2/9$ (see Rubinstein \& Clark \cite{rubinstein04}).

\subsection{Model constants}\label{DMC}

The closure model for $\bm{\mathcal{N}}+\bm{\mathcal{N}}^\dagger$ in (\ref{BHRZ_closure_N}), expressed through the terms $\bm{\mathcal{A}}$, $\bm{\mathcal{T}}^y$ and $\bm{\mathcal{T}}^k$, contains the unspecified, dimensionless constants $c_D$, $c_1$, $c_2$ and $c_M$. All but $c_D$, the turbulent diffusion coefficient, can be specified by considering theoretical, asymptotic constraints on the model for $\bm{E}(y,k,t)$. In particular, the expected scaling of $\bm{E}(y,k,t)$ for $k$ in the inertial range when $Re_\lambda\to\infty$, and the constraint that in the absence of transport in $k$-space, the $k$-space behavior must satisfy the equipartition spectrum $\bm{E}(y,k,t)\propto k^2$ (see Clark \& Zemach \cite{clark95} for more details on the use of these asymptotic constraints to determine $c_1$, $c_2$ and $c_M$). The turbulent diffusion coefficient $c_D$ cannot be specified using such asymptotic constraints because it is related to the physical-space transport, and this is dominated by the large scales of the flow which are entirely problem dependent. The only way to specify $c_D$ is to match it to known data, and the SFML is a particularly good test case for doing this since in the SFML the transport of energy in physical-space is entirely generated by the turbulence itself. 

It is worth mentioning that obtaining $c_D$ in this way does not in any way guarantee that the BHRZ model will accurately predict how the physical-space energy transport will vary across the mixing layer. This follows since $c_D$ does not control the functional behavior of $\bm{\mathcal{T}}^y$, it simply changes its magnitude. 

For the SFML, there are several ways that we could obtain an estimate for $c_D$, however, we choose to obtain the value of $c_D$ by determining the value that gives the best predictions from the BHRZ model when compared to the DNS data for the evolution of the mixing layer width $\Delta(t)\equiv Y_{[1/4]}-Y_{[3/4]}$, where $\mathscr{E}(Y_{[1/4]},t)=1/4$, $\mathscr{E}(Y_{[3/4]},t)=3/4$  and
\begin{align}
\mathscr{E}(y,t)&\equiv \frac{\min[\mathcal{K}(y,t)]-\mathcal{K}(y,t)}{\min[\mathcal{K}(y,t)]-\max[\mathcal{K}(y,t)]},
\end{align}
is a normalized energy function $\mathscr{E}(y,t)\in[0,1]$.

There are at least two reasons for choosing this approach to determine $c_D$: First, $\Delta(t)$ is a function of only one dimension (time), making it a much simpler candidate for determining $c_D$ than say $\mathcal{K}$, which depends on both $y$ and $t$. Second, in Tordella \& Iovieno \cite{tordella11} they found that $\Delta(t)$ has quite a weak dependence on both $Re_\lambda$ and $\max[\mathcal{K}(y,0)]/\min[\mathcal{K}(y,0)]$. Therefore, determining $c_D$ using $\Delta(t)$ will give a value which should depend weakly upon the initial conditions of the mixing layer. We choose to determine $c_D$, as described above, for the case with the smallest $\max[\mathcal{K}(y,0)]/\min[\mathcal{K}(y,0)]$. The reasoning behind this is that the turbulent transport closure employed in the BHRZ model is only strictly applicable for the case where the PDF of $\bm{u}(\bm{x},t)$ is weakly perturbed from a Gaussian. Such a situation would be realized in the limit $\max[\mathcal{K}(y,0)]/\min[\mathcal{K}(y,0)]\to1$ and it is therefore in this limit that the BHRZ closure model should be most appropriate. In \S\ref{ResDis} we will discuss the choice of $c_D$ further and the value we obtain for the SFML.

\subsection{Initial \& Boundary Conditions}

With $\langle\bm{u}(y,0)\bm{u}(y,0)\rangle$ specified in (\ref{uuinit}), we construct the initial condition $\bm{E}(\bm{x},k,0)$ (required in the BHRZ model) as
\begin{align}
\begin{split}
\bm{E}(y,k,0)&=\frac{\mathcal{E}(k,0)}{2\mathcal{K}^{[1]}}\Big\langle\bm{u}(y,0)\bm{u}(y,0)\Big\rangle=\frac{\mathcal{E}(k,0)}{3}\mathbf{I}\Big(P^2 +\gamma[1-P]\Big(2 P+\gamma[1-P]\Big)\Big),
\end{split}
\end{align}
$\mathcal{E}(k,0)$ being the energy spectrum corresponding to the homogeneous field $\bm{u}^{[1]}(\bm{x},0)$. We use the same data for the initial energy spectrum $\mathcal{E}(k,0)$ as was used in for the initial condition for the DNS in Tordella \emph{et al.} \cite{tordella08}. One difference is that in the DNS, $\mathcal{E}(k,0)$ is only prescribed for $k\in[1,k_{max}]$ whereas in the BHRZ model we want to solve for $\bm{E}(y,k,t)$ on the interval $k\in[k_{min},k_{max}]$, where $k_{min}\ll1$ (to ensure the validity of the boundary condition at $k_{min}$, discussed below) and $k_{max}$ is the maximum wavenumber. In order to specify $\mathcal{E}(k,0)$ for $k<1$ for the BHRZ model initial condition, we extrapolate the DNS data for $\mathcal{E}(k,0)$, ensuring that the infrared power-law scaling of the extrapolated data for $\mathcal{E}(k,0)$ for $k<1$ is consistent with the scaling of the original DNS data for $\mathcal{E}(k,0)$ at $k=\mathcal{O}(1)$.
 
We match the initial $Re_\lambda$ of the flow specified in the model (through the value specified for $\nu$) with that of the DNS. From the data for $\mathcal{E}(k,0)$ we can compute $u'_0$, the r.m.s. velocity of $\bm{u}^{[1]}(\bm{x},0)$ and then $\lambda_f$ (longitudinal Taylor microscale) through the relation \[{\lambda_f=\sqrt{2 u'_0u'_0/\langle[\nabla_1 u^{[1]}_{1}(y,t)]^2\rangle}}.\]We then calculate the appropriate value of $\nu$ to use in the BHRZ model through $\nu=u'_0\lambda_f/Re_\lambda$, and in Tordella \emph{et al.} \cite{tordella08} the Taylor Reynolds number of $\bm{u}^{[1]}(\bm{x},0)$ is $Re_\lambda=45$. Using the energy spectrum data from Tordella \emph{et al.} \cite{tordella08} we have $\max[\mathcal{K}(y,0)]=0.922$, $u'_0=\sqrt{(2/3)\max[\mathcal{K}(y,0)]}=0.784$ which gives $\nu\approx 0.0017$, and we use this value when solving the BHRZ model.

In Tordella \emph{et al.}  \cite{tordella08}, periodic boundary conditions were used for $\bm{u}(\bm{x},t)$, with period $L$ in the $y$ direction. Corresponding to this, we specify periodic boundary conditions in the $y$ domain, namely $\bm{E}(y,k,t)=\bm{E}(y+L,k,t)$. The boundary conditions in $k$-space are specified as follows: We introduce the variable $z\equiv \log[k]$, and enforce that the numerical grid be uniform in $z$-space. For small $k$, $\bm{E}(y,k,t)=\bm{J}k^n\implies\bm{E}(y,z,t)=\bm{J}e^{zn}$, where $\bm{J}$ is some tensor function of $y$ and $t$. Using a Taylor series we obtain the boundary condition for dummy node $z_{min}-\delta z$ (where $z_{min}\equiv\log[k_{min}]$)
\begin{align}
\begin{split}
\bm{E}(y,z_{min}-\delta z,t)&=\bm{E}(y,z_{min},t)-\delta z\partial_{z}\bm{E}(y,z_{min},t)+\mathcal{O}(\delta z^2)\\
&=\bm{J}e^{z_{min}n}-n \delta z \bm{J}e^{z_{min}n}+\mathcal{O}(\delta z^2)\\
&=e^{-n\delta z}\bm{E}(y,z_{min},t)+\mathcal{O}(\delta z^2),
\end{split}
\end{align}
and we take $n=2$, consistent with the behavior of the DNS data for $\mathcal{E}(k,0)$ at $k=\mathcal{O}(1)$.

The boundary condition for $k_{max}$ is specified as follows: Assuming that at high-$k$, $\bm{E}(y,k,t)=k^\xi\widetilde{\bm{E}}_{min}$, where $\widetilde{\bm{E}}_{min}\equiv k_{min}^{-1}\bm{E}(y,k_{min},t)$, then in $z$-space we have $\bm{E}(y,z,t)=\widetilde{\bm{E}}_{min}e^{z\xi}$. From this we construct the boundary condition for dummy node $z_{max}+\delta z$ ($z_{max}\equiv\log[k_{max}]$), which in component form is
\begin{align}
E_{ij}(y,z_{max}+\delta z,t)=E^2_{ij}(y,z_{max},t)\Big/E_{ij}(y,z_{max}-\delta z,t).
\end{align}

\section{Results \& Discussion}\label{ResDis}

\begin{figure}[ht]
\centering
\vspace{-7mm}
\subfloat[]
{\begin{overpic}
[trim = 0mm 50mm 0mm 60mm,scale=0.4,clip,tics=20]{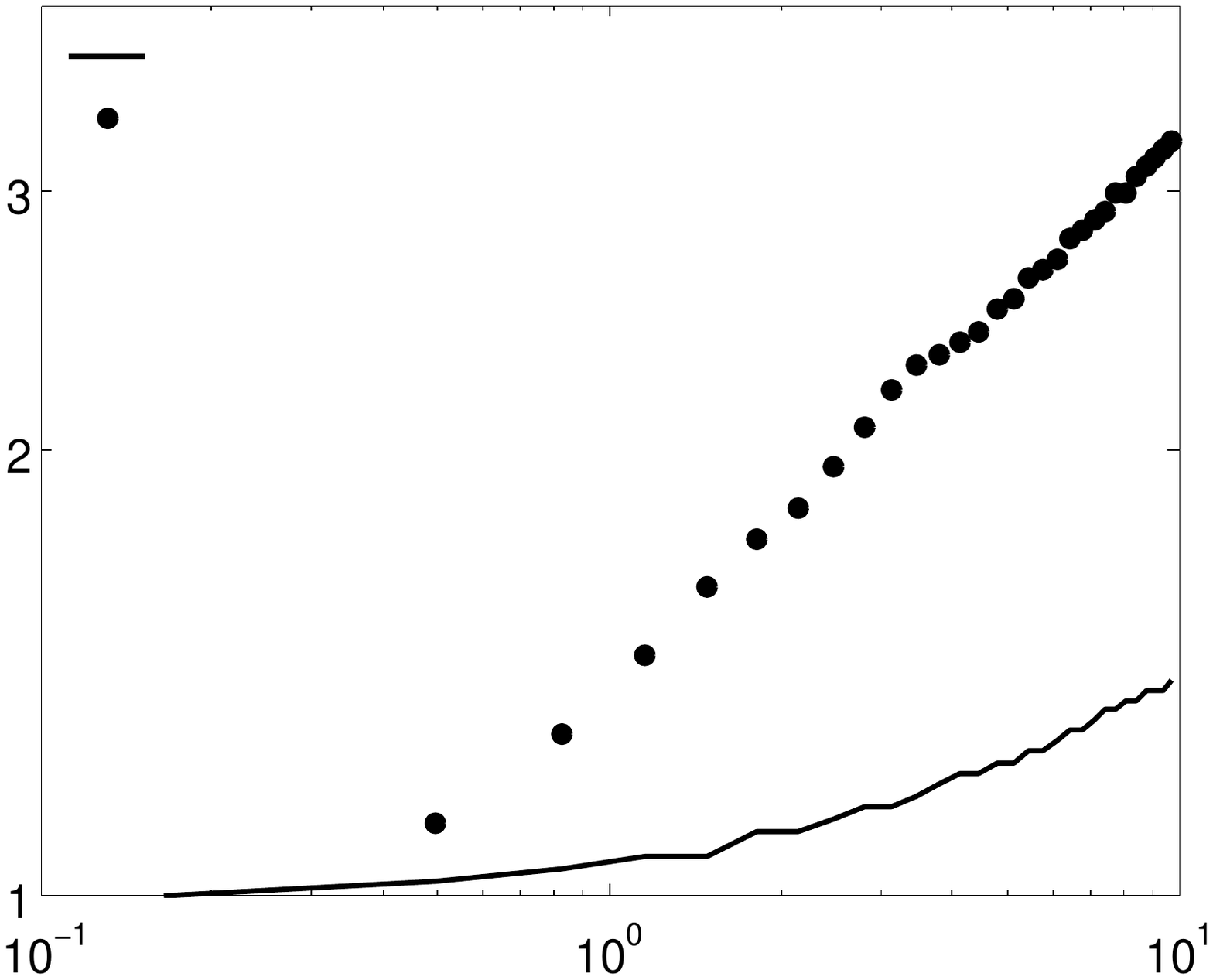}
\put(119,12){$t/\tau$}
\put(15,80){\rotatebox{90}{$\Delta(t)/\Delta(0)$}}
\put(55,164){\tiny{BHRZ $\gamma=\sqrt{1/40}$}}
\put(55,154){\tiny{DNS \hspace{2mm}$\gamma=\sqrt{1/40}$}}
\end{overpic}}
\subfloat[]
{\begin{overpic}
[trim = 0mm 50mm 0mm 60mm,scale=0.4,clip,tics=20]{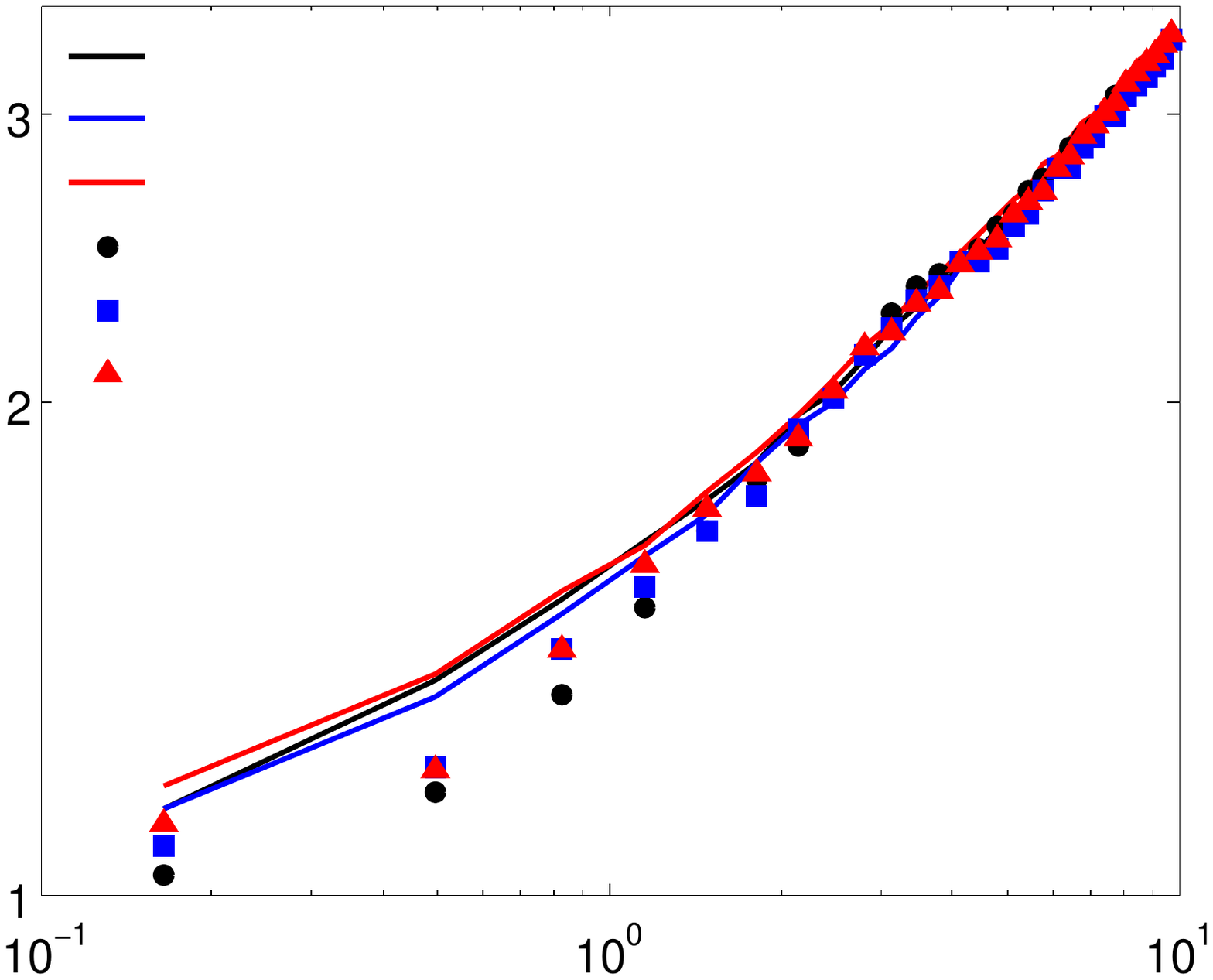}
\put(120,15){$t/\tau$}
\put(15,80){\rotatebox{90}{$\Delta(t)/\Delta(0)$}}
\put(55,164){\tiny{BHRZ $\gamma=\sqrt{1/40}$}}
\put(78,154){\tiny{$\gamma=\sqrt{1/100}$}}
\put(78,144){\tiny{$\gamma=\sqrt{1/300}$}}
\put(55,135){\tiny{DNS \hspace{2mm}$\gamma=\sqrt{1/40}$}}
\put(78,125){\tiny{$\gamma=\sqrt{1/100}$}}
\put(78,115){\tiny{$\gamma=\sqrt{1/300}$}}
\end{overpic}}
\caption{Plot of BHRZ predictions and DNS data for $\Delta(t)/\Delta(0)$ with (a) $c_D=0$ and (b) $c_D=0.025$. Black line/symbols are for $\gamma=\sqrt{1/40}$, blue line/symbols are for $\gamma=\sqrt{1/100}$ and red line/symbols are for $\gamma=\sqrt{1/300}$.}
\label{MLW_40}
\end{figure}

We begin by considering the results for the mixing layer width $\Delta(t)$, introduced in \S\ref{DMC}. In \S\ref{DMC} we discussed that the BHRZ value for $c_D$ would be chosen as that value which gives the best fit for the BHRZ solution for $\Delta(t)$ compared with the DNS data. In Figure~\ref{MLW_40}(a) we first compare the DNS data with the BHRZ prediction obtained with $c_D=0$, in order to determine how much of a contribution the turbulent transport gives to $\Delta(t)$ compared with the molecular diffusion contribution. The results indicate that the turbulence transport is by far the most dominant contribution to the evolution of $\Delta(t)$, even at the relatively low $Re_\lambda$ of the flow ($\approx 45$ at $t=0$).

For the case with $\gamma=\sqrt{1/40}$ (weakest initial inhomogeneity for which we have DNS data) we determined the value $c_D=0.025$. From Figure~\ref{MLW_40}(b) it can be seen that this choice of $c_D$ gives good agreement between BHRZ the DNS data at long times, both qualitatively and quantitatively. Although another value could have been chosen to give a good match at small times, this would have led to significant errors at large times. Furthermore, on physical grounds it makes more sense to match the solutions at longer times since the turbulence inhomogeneity weakens with increasing time, and it is in the regime of weak inhomogeneity that the approximations invoked in the BHRZ are most justified. 

The value of $c_D$ that we obtain is quite different to the value obtained for one-point models, e.g. in Hanjalic \& Launder \cite{hanjalic72} the turbulent transport coefficient has a value $0.11$. However, there is no reason to expect $c_D$ to coincide with that in one-point models since the diffusion tensor in one-point models, $\mathbf{D}(y,t)$, is often defined as $\mathbf{D}(y,t)\equiv (\mathcal{K}/\langle\epsilon\rangle)\bm{R}_2$ and this is not equal to the diffusion tensor $\bm{\mathscr{D}}(y,t)$ in the BHRZ model.

The results in Fig.~\ref{MLW_40}(a) reveal that for $t/\tau<2$ the BHRZ predictions are in error, overpredicting $\Delta(t)$ compared with the DNS. There are at least three possible reasons why $\bm{\mathcal{T}}^y$ could lead to these errors: First, the local approximation to the intrinsically non-local pressure-transport, second, the gradient-diffusion approximation for $\bm{\mathcal{R}}_3$ and third, the neglect of certain transient effects present in the DNS but not present in the model described by $\bm{\mathcal{T}}^y$. Considering the first two, although these are surely sources of error in the SFML where the inhomogeneity is strong, it is difficult to say what the nature of the errors introduced by these approximations would be, i.e. whether they would lead to over or under predictions of the energy transport in $y$-space. Regarding the third reason, since the initial field in the DNS is constructed from weighted contributions of the homogeneous, isotropic velocity fields $\bm{u}^{[1]}(\bm{x},0)$ and $\bm{u}^{[2]}(\bm{x},0)$, the initial velocity field in the SFML satisfies $\langle\bm{u}(y,0)\bm{u}(y,0)\bm{u}(y,0)\rangle=\bm{0}$. Consequently, in the DNS, at $t=0$ there is no transport of energy in physical-space. However, in the BHRZ model, $\bm{\mathcal{T}}^y(y,k,0)\neq\bm{0}$, which would lead to an overprediction of the initial growth of $\Delta(t)$. To examine whether or not this is the cause of the discrepancy observed in Fig.~\ref{MLW_40}(a) we performed the following test case. We used the DNS data at $t/\tau=1$ as the `initial condition' for the BHRZ model and then ran the model. In this case, at the `initial time' $t/\tau=1$, both the DNS and BHRZ model have a velocity field with finite energy transport in physical-space. The results showed little difference \footnote[1]{This test case also rules out another possible cause of the discrepancies: In the DNS, the initial velocity field is somewhat `artificial' and it will take some time before the velocity field generated in the SFML becomes a true Navier-Stokes solution, with the correct coupling between the scales of motion etc. We would expect that this transition to the true Navier-Stokes dynamics takes $\mathcal{O}(\tau)$ to occur. Since the test case used  $t/\tau=1$, as the `initial condition', the fact that the results showed little difference with those obtained using the true initial condition, $t/\tau =0$, suggest that the effect of the artificiality of the initial condition in the DNS is only playing a minor role for the physical mechanisms governing the mixing in the velocity field.}, suggesting that the overpredictions in Fig.~\ref{MLW_40}(a) for $t/\tau<2$ are in fact caused by either the local approximation to the pressure-transport or by the gradient diffusion model for $\bm{\mathcal{R}}_3$. 
\FloatBarrier
We would argue however that the local approximation to the pressure-transport is the main source of error. The reason for this is that in the initial stage of the evolution of the SFML, the  PDF of $\bm{u}(\bm{x},t)$ is only slightly perturbed from its initial Gaussian form, and in that case the gradient-diffusion approximation for $\bm{\mathcal{R}}_3$ should be quite accurate. In contrast, the local approximation to the pressure-transport will be in greatest error during the initial stage of the evolution of the SFML since the inhomogeneity of the flow weakens with increasing $t$.

\begin{figure}[ht]
\centering
\vspace{-10mm}
\subfloat[]
{\begin{overpic}
[trim = 0mm 60mm 0mm 60mm,scale=0.4,clip,tics=20]{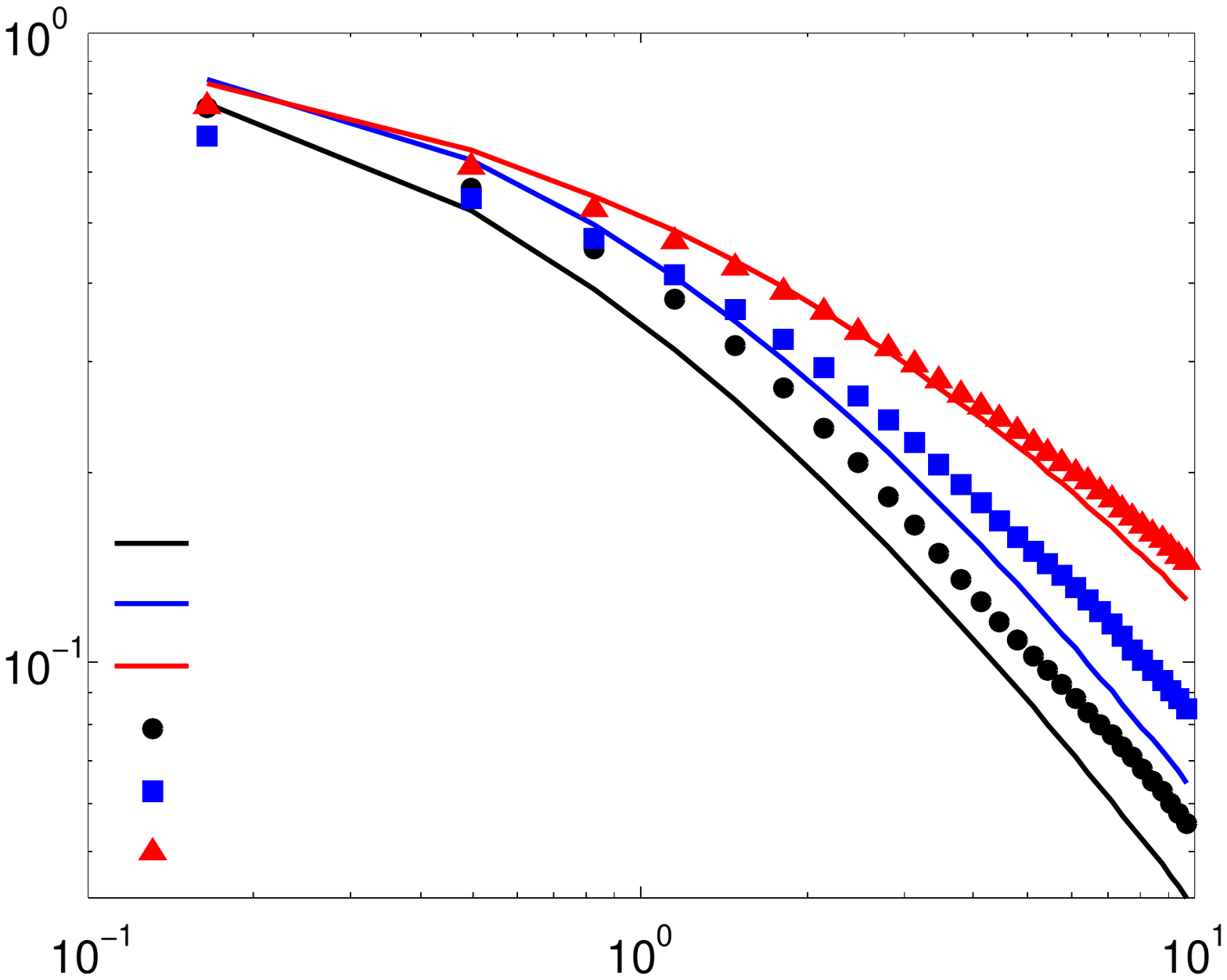}
\put(117,3){$t/\tau$}
\put(10,65){\rotatebox{90}{$\mathcal{K}(y,t)/\mathcal{K}(y,0)$}}
\put(55,79){\tiny{BHRZ $y-y_c=-2$}}
\put(78,69){\tiny{$y-y_c=0$}}
\put(78,60){\tiny{$y-y_c=+2$}}
\put(55,50){\tiny{DNS \hspace{2mm}$y-y_c=-2$}}
\put(78,40){\tiny{$y-y_c=0$}}
\put(78,30){\tiny{$y-y_c=+2$}}
\end{overpic}}
\subfloat[]
{\begin{overpic}
[trim = 0mm 60mm 0mm 60mm,scale=0.4,clip,tics=20]{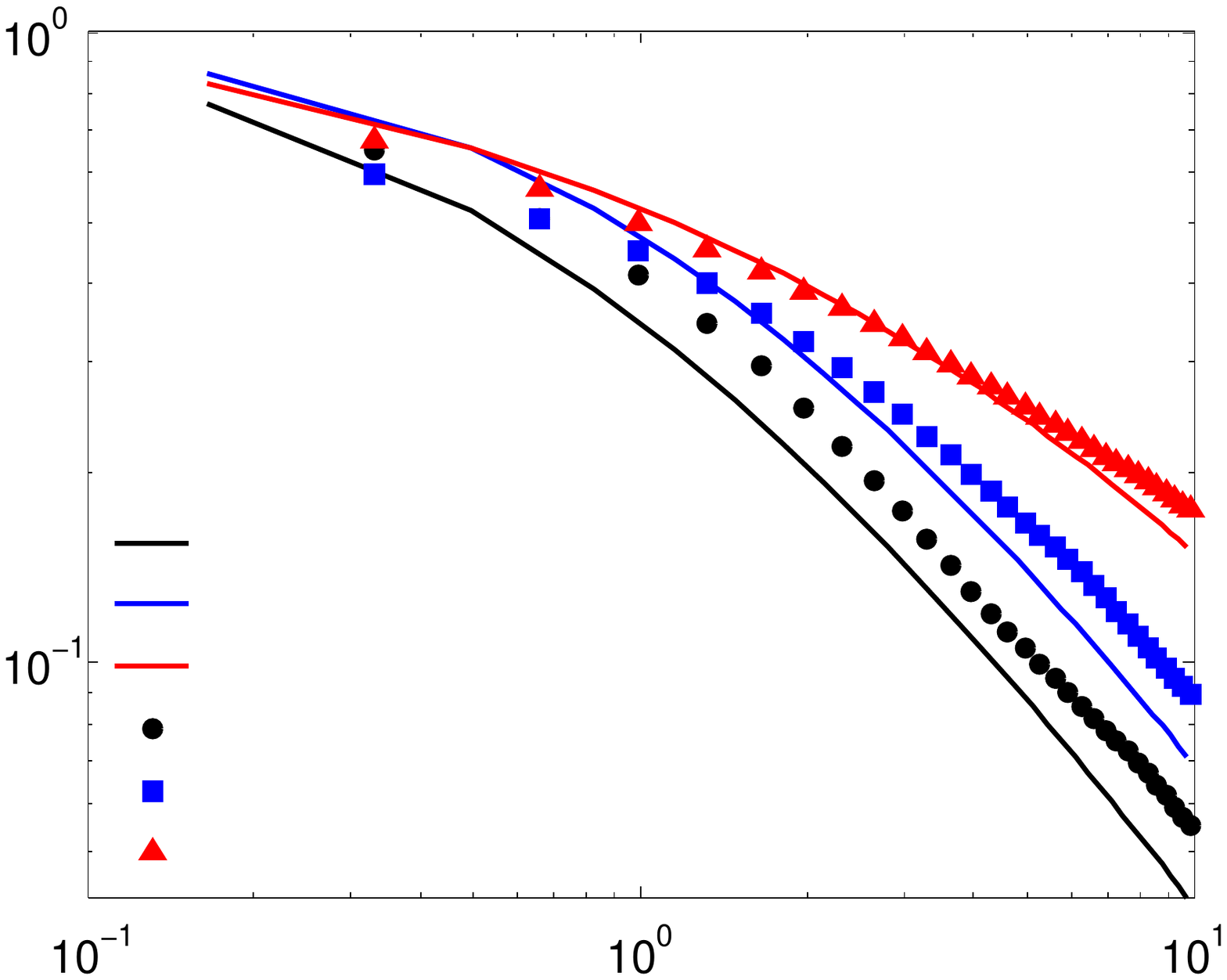}
\put(117,3){$t/\tau$}
\put(10,65){\rotatebox{90}{$\mathcal{K}(y,t)/\mathcal{K}(y,0)$}}
\put(55,79){\tiny{BHRZ $y-y_c=-2$}}
\put(78,69){\tiny{$y-y_c=0$}}
\put(78,60){\tiny{$y-y_c=+2$}}
\put(55,50){\tiny{DNS \hspace{2mm}$y-y_c=-2$}}
\put(78,40){\tiny{$y-y_c=0$}}
\put(78,30){\tiny{$y-y_c=+2$}}
\end{overpic}}
\caption{Plot of BHRZ predictions and DNS data for $\mathcal{K}(y,t)/\mathcal{K}(y,0)$ at various $y$ for (a) $\gamma=\sqrt{1/40}$ and (b) $\gamma=\sqrt{1/300}$. Black symbols/line correspond to $y-y_c=-2$, blue symbols/line to $y-y_c=0$ and red symbols/line to $y-y_c=+2$.}
\label{Edecay}
\end{figure}

We now consider the BHRZ predictions and DNS data for $\mathscr{E}(y,t)$, which, unlike $\Delta(t)$, not only gives a measure of the mixing with time, but also how the energy behaves as a function of position. The results show that the BHRZ model is able to predict $\mathscr{E}(y,t)$ quite well, capturing dependence on $y$\footnote[2]{See appendix for an explanation for the discrepancies at $y-y_c\lesssim-0.4$ for $t/\tau=6.765$.}. That the BHRZ model predicts the dependence of $\mathscr{E}(y,t)$ on $y$ as well as it does is somewhat surprising given that the transport model described by $\bm{\mathcal{T}}^y$ is only strictly appropriate in the limit where the  PDF of $\bm{u}(\bm{x},t)$ is weakly perturbed from a Gaussian; the results in Tordella \emph{et al.} \cite{tordella08} show that even for $\gamma=\sqrt{1/40}$, the  PDF of $\bm{u}(\bm{x},t)$ is quite far from being Gaussian. This suggests one of two things: First, it could simply suggest that the contributions from higher order cumulants of the field $\bm{u}(\bm{x},t)$ to the physical-space transport are small. Second, it could imply that the contributions from higher order cumulants of the field $\bm{u}(\bm{x},t)$ to the physical-space transport generate a similar dependance on $y$ across the mixing layer (such that their effect is subsumed in the value determined for $c_D$). 
\begin{figure}[ht]
\centering
\vspace{-7mm}
\subfloat[]
{\begin{overpic}
[trim = 0mm 60mm 0mm 60mm,scale=0.4,clip,tics=20]{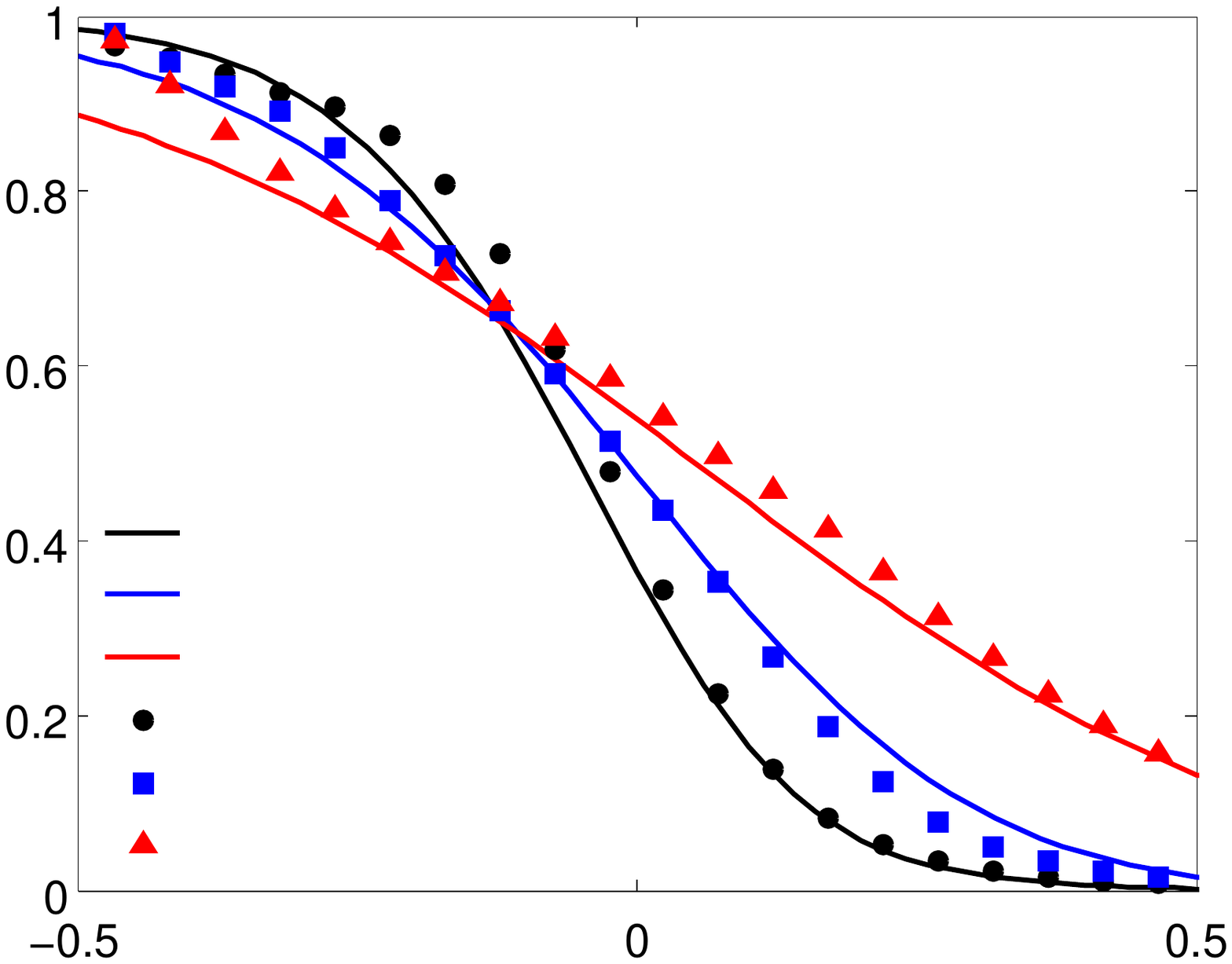}
\put(113,5){$y-y_c$}
\put(10,75){\rotatebox{90}{$\mathscr{E}(y,t)$}}
\put(55,79){\tiny{BHRZ $t/\tau=0.165$}}
\put(78,69){\tiny{$t/\tau=1.485$}}
\put(78,60){\tiny{$t/\tau=6.765$}}
\put(55,50){\tiny{DNS \hspace{2mm}$t/\tau=0.165$}}
\put(78,40){\tiny{$t/\tau=1.485$}}
\put(78,30){\tiny{$t/\tau=6.765$}}
\end{overpic}}
\subfloat[]
{\begin{overpic}
[trim = 0mm 60mm 0mm 60mm,scale=0.4,clip,tics=20]{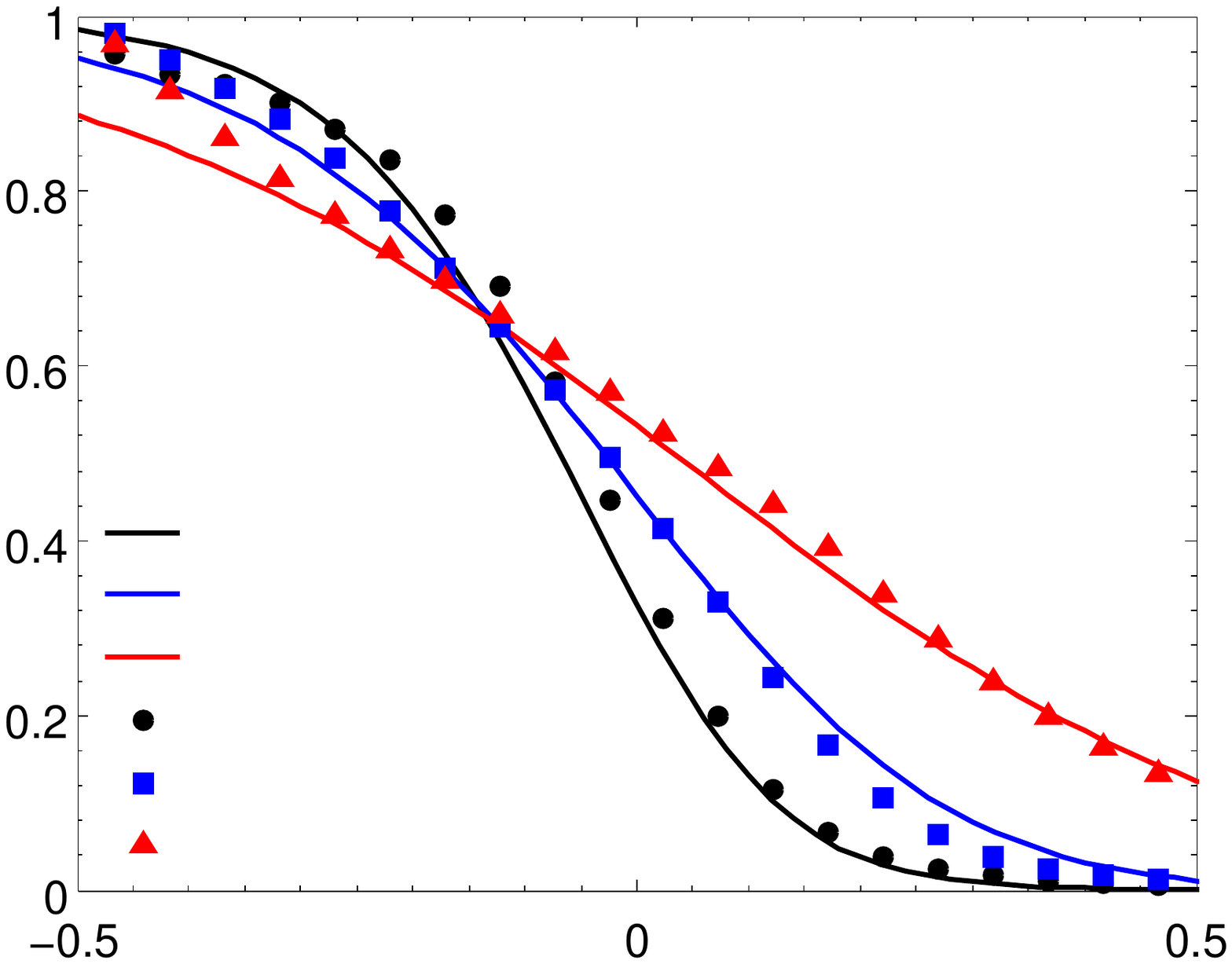}
\put(113,5){$y-y_c$}
\put(10,75){\rotatebox{90}{$\mathscr{E}(y,t)$}}
\put(55,79){\tiny{BHRZ $t/\tau=0.165$}}
\put(78,69){\tiny{$t/\tau=1.485$}}
\put(78,60){\tiny{$t/\tau=6.765$}}
\put(55,50){\tiny{DNS \hspace{2mm}$t/\tau=0.165$}}
\put(78,40){\tiny{$t/\tau=1.485$}}
\put(78,30){\tiny{$t/\tau=6.765$}}
\end{overpic}}
\caption{Plot of BHRZ predictions (lines) and DNS data (symbols) for $\mathscr{E}(y,t)$ as a function of $y-y_c$ (where $y_c=L/2$) for (a) $\gamma=\sqrt{1/40}$ and (b) $\gamma=\sqrt{1/300}$. Black symbols/line correspond to $t/\tau=0.165$, blue symbols/line to $t/\tau=1.485$ and red symbols/line to $t/\tau=6.765$.}
\label{Energy_func}
\end{figure}
\FloatBarrier
Next, we compare the BHRZ predictions and the DNS data for the evolution of the turbulent kinetic energy $\mathcal{K}(y,t)/\mathcal{K}(y,0)$. In an inhomogeneous turbulent velocity field there are essentially two ways that $\mathcal{K}(y,t)/\mathcal{K}(y,0)$ can change; through transport of energy to/away from $y$, and through viscous dissipation. For $y-y_c>0$, $\mathcal{K}(y,t)/\mathcal{K}(y,0)$ may behave non-monotonically since these regions are receiving energy from the high energy side of the mixing layer, which could cause $\mathcal{K}(y,t)/\mathcal{K}(y,0)$ to increase, whereas viscous dissipation effects cause $\mathcal{K}(y,t)/\mathcal{K}(y,0)$ to decrease. The results in Figure~\ref{Edecay} reveal that at this low $Re_\lambda$, viscous effects dominate and $\mathcal{K}(y,t)/\mathcal{K}(y,0)$ monotonically decreases. The BHRZ results are in reasonable agreement with the DNS, though with a consistent under-prediction in the long time regime. It is difficult to isolate a particular cause of these under-predictions since there are so many different effects in the system. One possibility is related to $\Phi$, which is based upon a dimensional estimate for the turbulence time scales, and may be therefore in error (relative to the `true' timescale) by some $\mathcal{O}(1)$ factor. In particular, whereas the mixing is dominated by the largest flow scales, the dissipation is dominated by the smallest scales, but in the BHRZ model, the rate at which the energy is transferred from the large to the small scales is dependent upon $\Phi$. Figure~\ref{Edecay} also shows that the BHRZ model gives a reasonably good match with the DNS data for the decay exponent of the turbulent kinetic energy at long times.

Having considered the BHRZ results for the behavior of the turbulent kinetic energy during the mixing process, we now consider the behavior of the particular components of $\bm{R}_2(y,t)$. Comparing the components of this tensor reveal the evolution of the velocity field anisotropy during the mixing process. In Tordella \emph{et al.} \cite{tordella08} it was observed that the velocity field evolved from its initially isotropic state to a self-similar anisotropic state for $t/\tau\gtrsim 2$, during which the anisotropy measure $R_{2,22}/\mathrm{tr}[\bm{R}_2]$ remained approximately constant. This self-similar state can be understood as the state where the anisotropy generated by the physical-space transport, and the return-to-isotropy effects approximately balance each other.

\begin{figure}[ht]
\centering
\vspace{-15mm}
\subfloat[]
{\begin{overpic}
[trim = 0mm 50mm 0mm 60mm,scale=0.4,clip,tics=20]{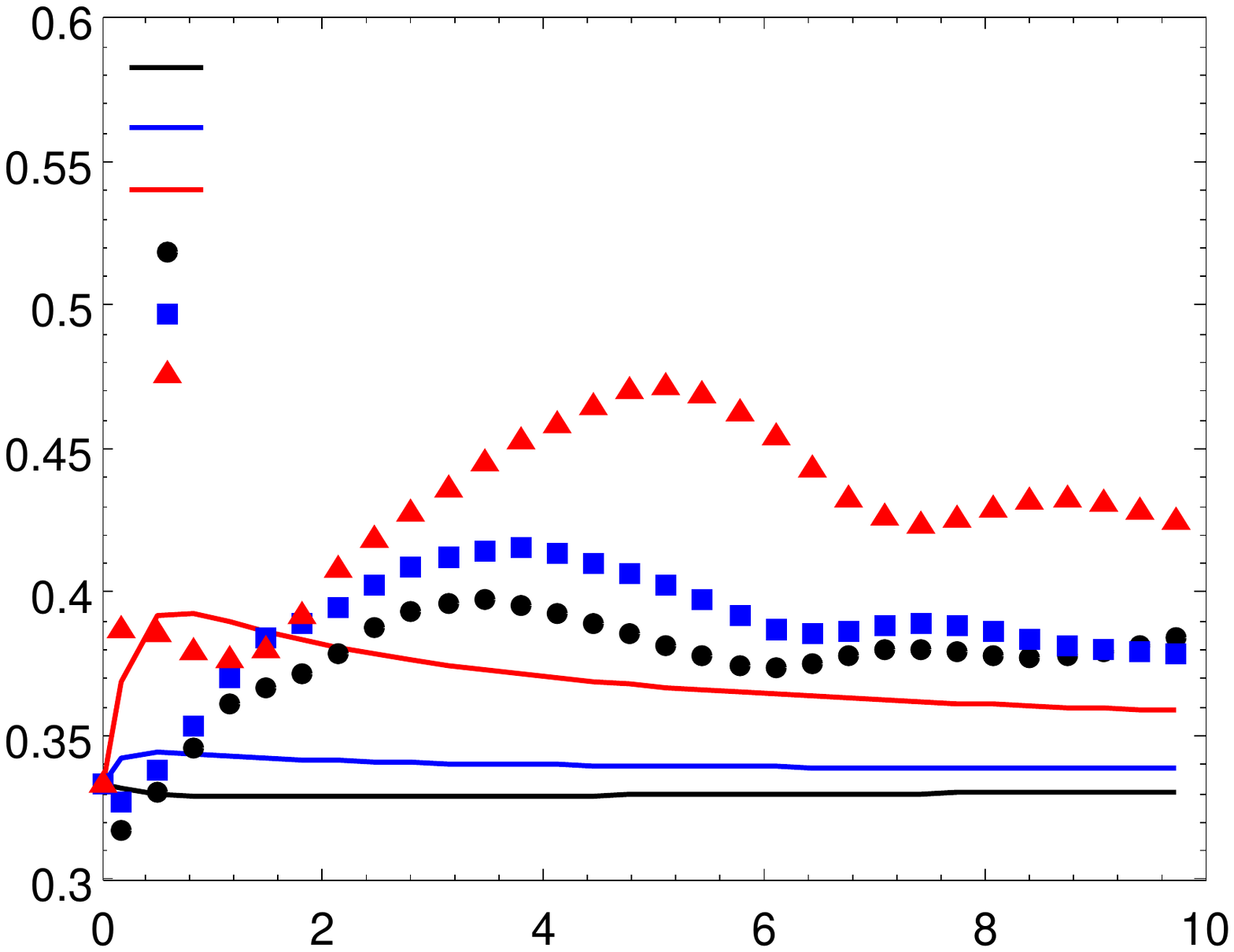}
\put(117,13){$t/\tau$}
\put(10,75){\rotatebox{90}{$R_{2,22}/\mathrm{tr}[\bm{R}_2]$}}
\put(55,164){\tiny{BHRZ $y-y_c=-2$}}
\put(78,154){\tiny{$y-y_c=0$}}
\put(78,145){\tiny{$y-y_c=+2$}}
\put(55,135){\tiny{DNS \hspace{2mm}$y-y_c=-2$}}
\put(78,125){\tiny{$y-y_c=0$}}
\put(78,115){\tiny{$y-y_c=+2$}}
\end{overpic}}
\subfloat[]
{\begin{overpic}
[trim = 0mm 50mm 0mm 60mm,scale=0.4,clip,tics=20]{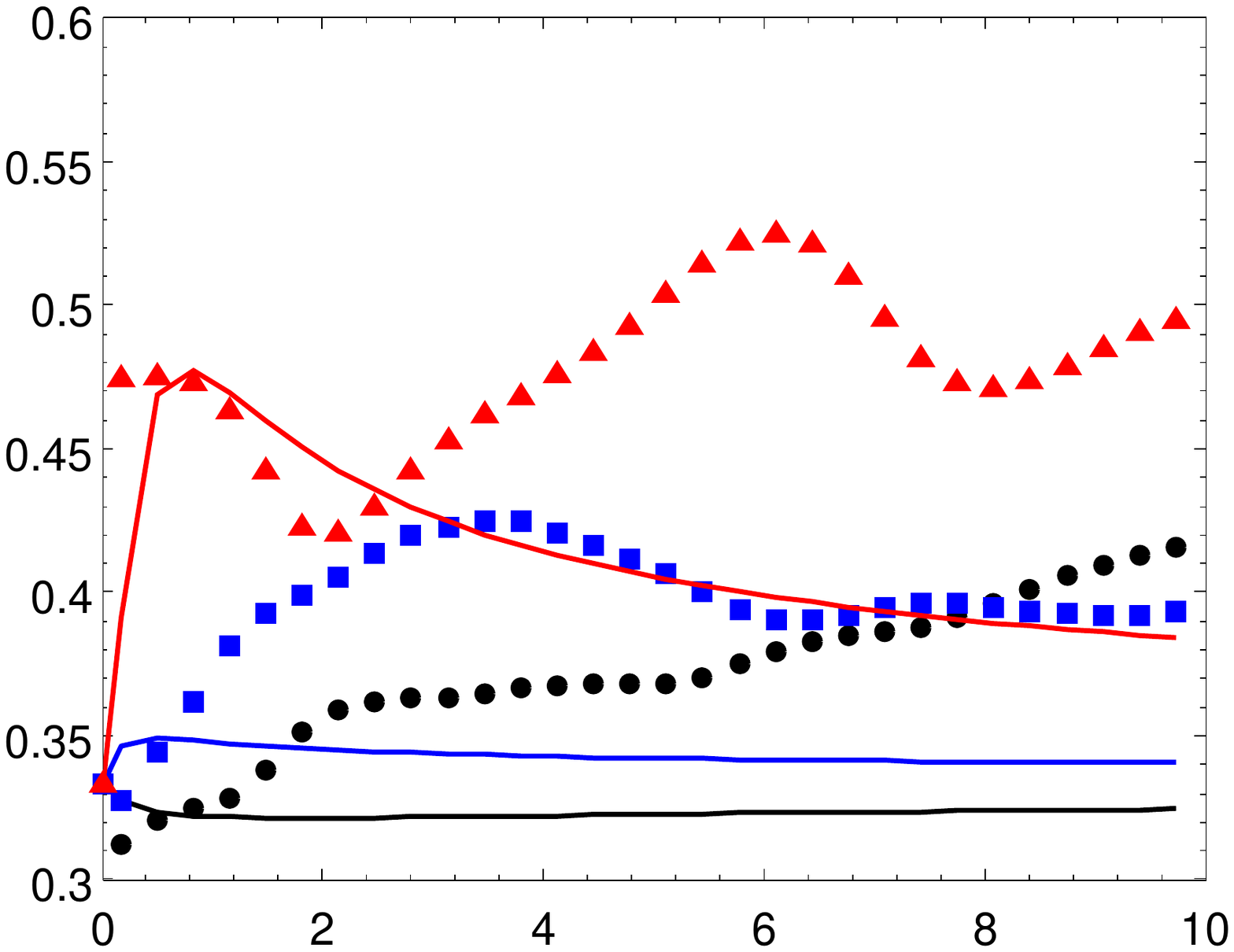}
\put(117,13){$t/\tau$}
\put(10,75){\rotatebox{90}{$R_{2,22}/\mathrm{tr}[\bm{R}_2]$}}
\end{overpic}}
\caption{Plot of BHRZ predictions and DNS data for $R_{2,22}/\mathrm{tr}[\bm{R}_2]$ at various $y-y_c$ for (a) $\gamma=\sqrt{1/40}$ and (b) $\gamma=\sqrt{1/300}$. Black symbols/line correspond to $y-y_c=-2$, blue symbols/line to $y-y_c=0$ and red symbols/line to $y-y_c=+2$. Plot (b) legend corresponds to that in plot (a).}
\label{Anis}
\end{figure}

\begin{figure}[ht]
\centering
\vspace{-2mm}
\subfloat[]
{\begin{overpic}
[trim = 0mm 50mm 0mm 60mm,scale=0.4,clip,tics=20]{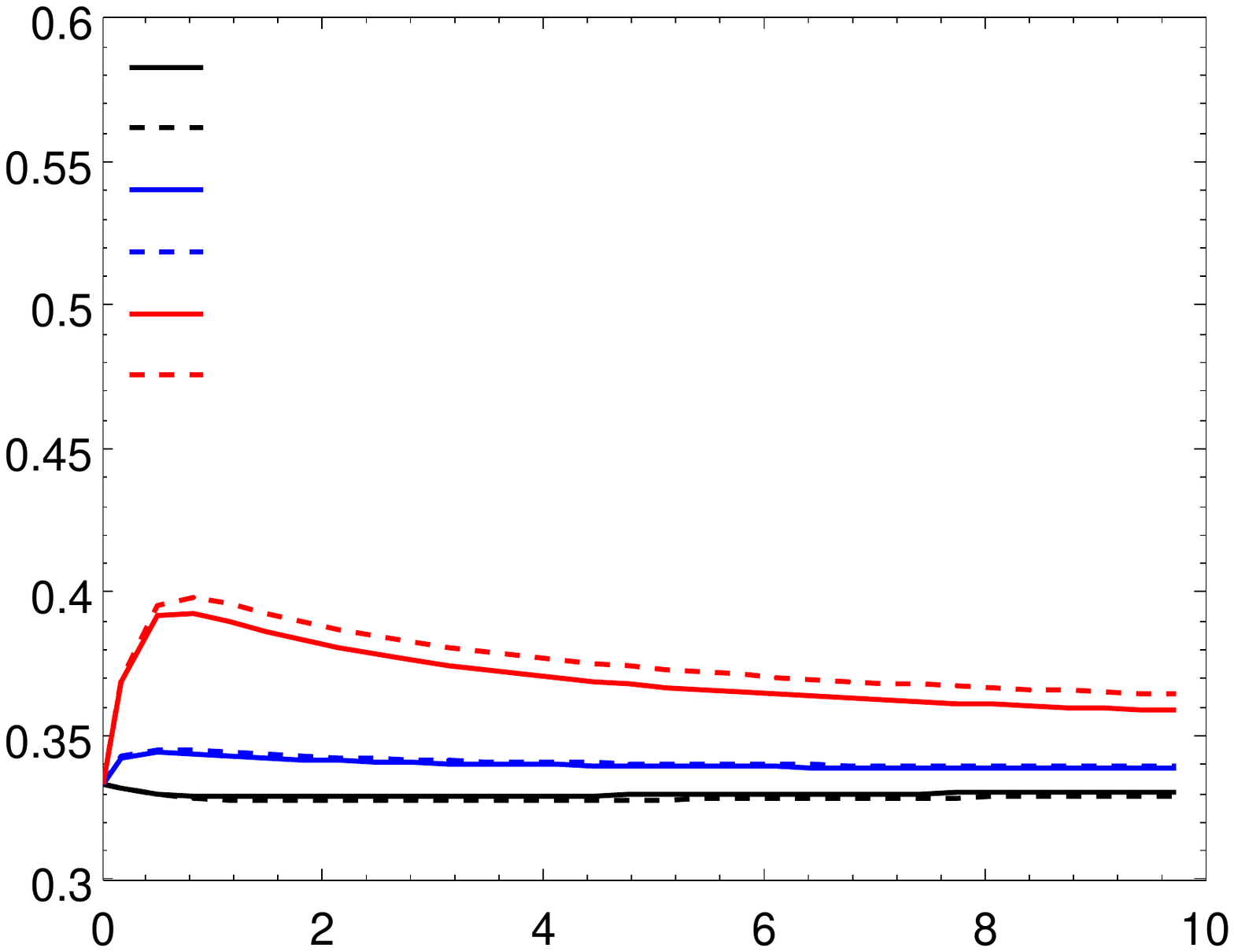}
\put(117,13){$t/\tau$}
\put(10,75){\rotatebox{90}{$R_{2,22}/\mathrm{tr}[\bm{R}_2]$}}
%\put(55,164){\scriptsize{BHRZ (with $c_M\neq0$)}}
%\put(55,154){\scriptsize{BHRZ (with $c_M=0$)}}
%\put(55,144){\scriptsize{BHRZ (with $c_M\neq0$)}}
%\put(55,134){\scriptsize{BHRZ (with $c_M=0$)}}
%\put(55,124){\scriptsize{BHRZ (with $c_M\neq0$)}}
%\put(55,115){\scriptsize{BHRZ (with $c_M=0$)}}
\put(55,164){\tiny{BHRZ $y-y_c=-2$, $c_M\neq0$}}
\put(78,154){\tiny{$y-y_c=0$, \hspace{1mm} $c_M=0$ }}
\put(78,145){\tiny{$y-y_c=+2$, $c_M\neq0$}}
\put(78,135){\tiny{$y-y_c=-2$, $c_M=0$}}
\put(78,125){\tiny{$y-y_c=0$, \hspace{1mm} $c_M\neq0$}}
\put(78,115){\tiny{$y-y_c=+2$, $c_M=0$}}
\end{overpic}}
\subfloat[]
{\begin{overpic}
[trim = 0mm 50mm 0mm 60mm,scale=0.4,clip,tics=20]{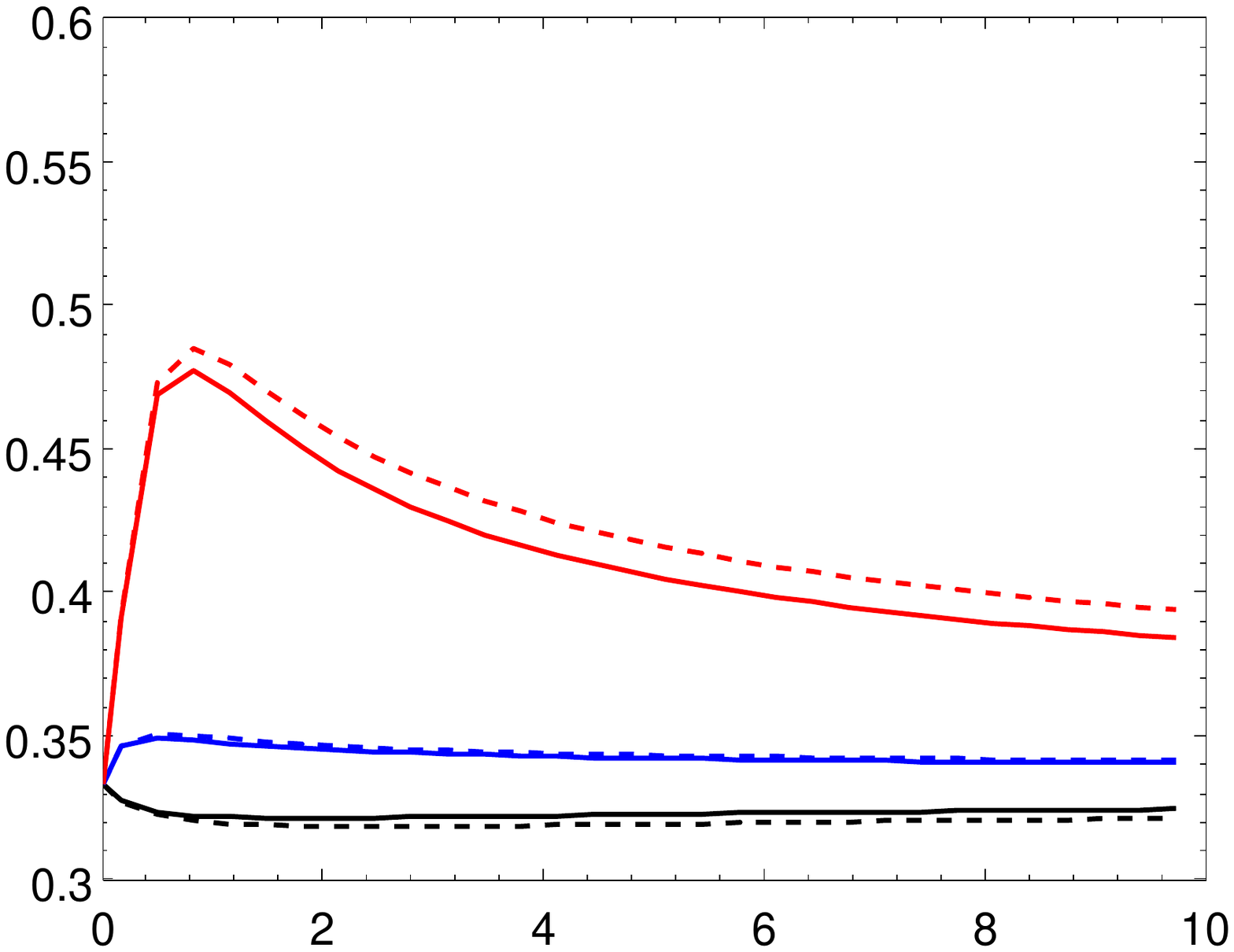}
\put(117,13){$t/\tau$}
\put(10,75){\rotatebox{90}{$R_{2,22}/\mathrm{tr}[\bm{R}_2]$}}
\end{overpic}}
\caption{Plot of BHRZ predictions and DNS data for $R_{2,22}/\mathrm{tr}[\bm{R}_2]$ at various $y-y_c$ for (a) $\gamma=\sqrt{1/40}$ and (b) $\gamma=\sqrt{1/300}$. Black lines correspond to $y-y_c=-2$, blue lines to $y-y_c=0$ and red lines to $y-y_c=+2$. Solid lines are the results using the original BHRZ value for $c_M$, while the dashed lines are the results using $c_M=0$. Plot (b) legend corresponds to that in plot (a).}
\label{Anis_cM_zero}
\end{figure}

\begin{figure}[ht]
\centering
\vspace{-2mm}
\subfloat[]
{\begin{overpic}
[trim = 0mm 60mm 0mm 60mm,scale=0.4,clip,tics=20]{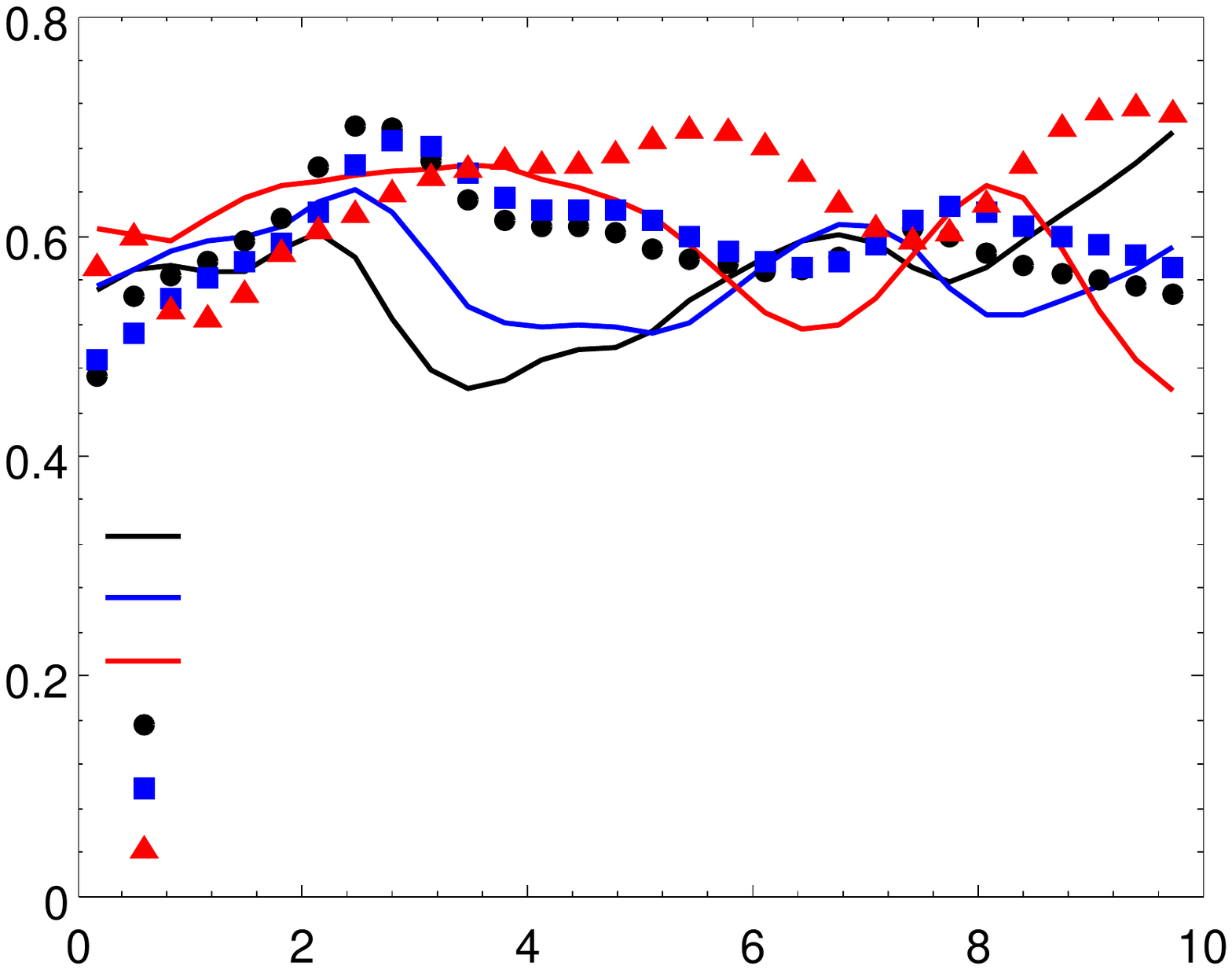}
\put(117,3){$t/\tau$}
\put(10,20){\rotatebox{90}{$R_{3,222}/(R_{3,112}+R_{3,222}+R_{3,332})$}}
\put(55,79){\tiny{Eq.(\ref{R3graddiff_2}) $y-y_c=-2$}}
\put(78,69){\tiny{$y-y_c=0$}}
\put(78,60){\tiny{$y-y_c=+2$}}
\put(55,50){\tiny{DNS \hspace{2mm}$y-y_c=-2$}}
\put(78,40){\tiny{$y-y_c=0$}}
\put(78,30){\tiny{$y-y_c=+2$}}
\end{overpic}}
\subfloat[]
{\begin{overpic}
[trim = 0mm 60mm 0mm 60mm,scale=0.4,clip,tics=20]{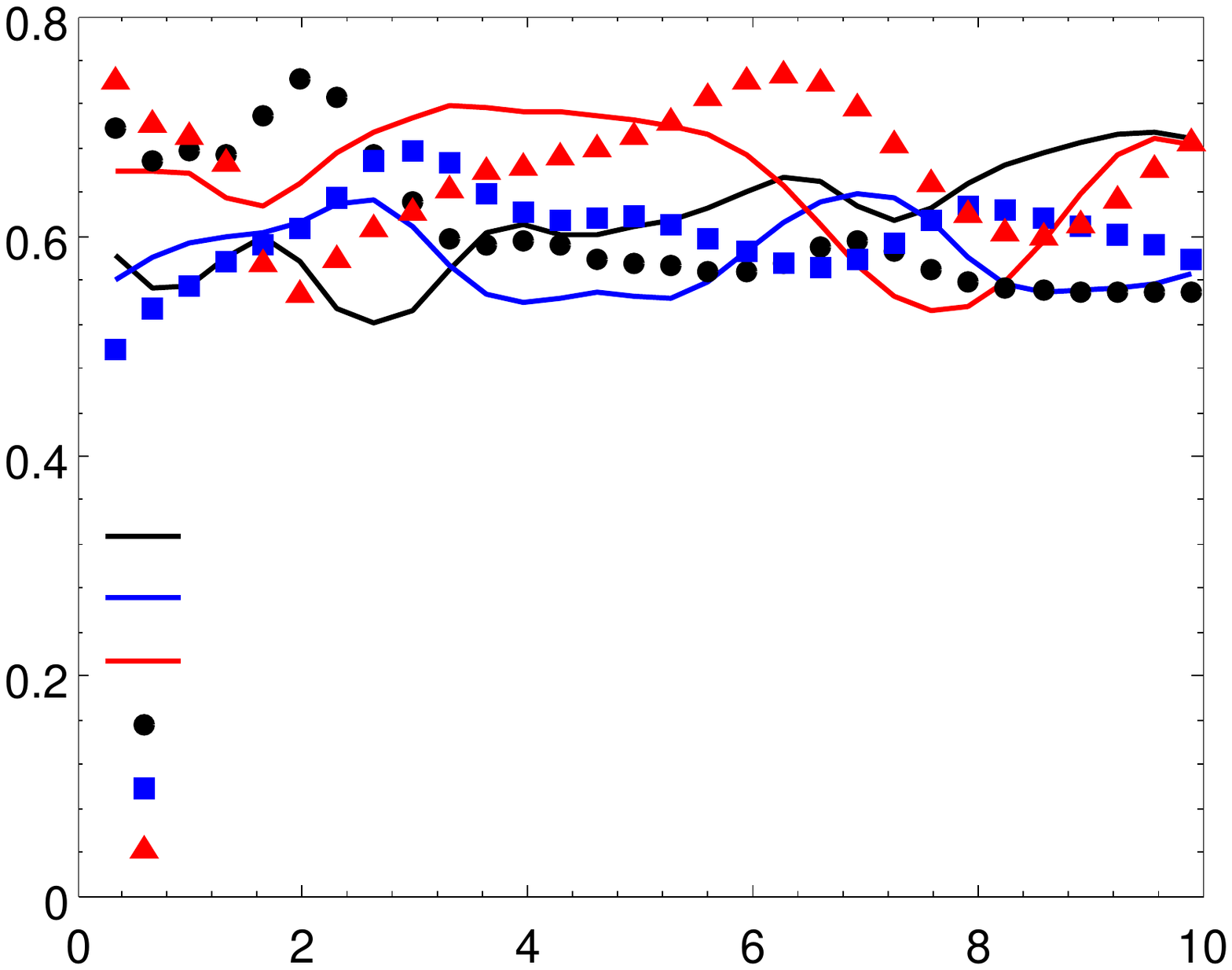}
\put(117,3){$t/\tau$}
\put(10,20){\rotatebox{90}{$R_{3,222}/(R_{3,112}+R_{3,222}+R_{3,332})$}}
\put(55,79){\tiny{Eq.(\ref{R3graddiff_2}) $y-y_c=-2$}}
\put(78,69){\tiny{$y-y_c=0$}}
\put(78,60){\tiny{$y-y_c=+2$}}
\put(55,50){\tiny{DNS \hspace{2mm}$y-y_c=-2$}}
\put(78,40){\tiny{$y-y_c=0$}}
\put(78,30){\tiny{$y-y_c=+2$}}
\end{overpic}}
\caption{Plot of predictions from (\ref{R3graddiff_2}) and DNS data for $R_{3,222}/(R_{3,112}+R_{3,222}+R_{3,332})$ at various $y-y_c$ for (a) $\gamma=\sqrt{1/40}$ and (b) $\gamma=\sqrt{1/300}$. Black symbols/line correspond to $y-y_c=-2$, blue symbols/line to $y-y_c=0$ and red symbols/line to $y-y_c=+2$. }
\label{R3_ratio}
\end{figure}

Figure~\ref{Anis} shows a comparison of the BHRZ predictions and DNS data for the anisotropy measure $R_{2,22}/\mathrm{tr}[\bm{R}_2]$. The results reveal that the BHRZ model is in significant error in describing the evolution of the flow field anisotropy, and in particular, it significantly under-predicts the anisotropy when the SFML is in its self-similar regime. Within the BHRZ modeling framework, the anisotropy is generated exclusively by $\bm{\mathcal{T}}^y$, and the term $\bm{\mathcal{A}}$ drives the system back towards isotropy. The under-predictions could therefore be because $\bm{\mathcal{T}}^y$ is too small or else because $\bm{\mathcal{A}}$ is too large. To consider this, in Fig.~\ref{Anis_cM_zero} we show results where we compare the BHRZ predictions using the original value of $c_M$ and also using $c_M=0$ (corresponding to $\bm{\mathcal{A}}=\bm{0}$). 
\FloatBarrier
\FloatBarrier
The results show reveal that for the BHRZ model, $\bm{\mathcal{A}}$ is playing a small role and that the behavior of $R_{2,22}/\mathrm{tr}[\bm{R}_2]$ predicted by BHRZ is almost entirely controlled by $\bm{\mathcal{T}}^y$. As discussed earlier, these errors are either associated with the angle averaging operation applied to the transport equation, the local approximation to the pressure-transport or else the gradient-diffusion closure for $\bm{\mathcal{R}}_3$. Although it is difficult to estimate the effect of the first two approximations, we can obtain an estimate of the accuracy of the latter approximation.
\FloatBarrier
The physical-space transport of $\bm{R}_2$ arising from the nonlinear advection and pressure gradient terms in the NSE is given by $\bm{\mathscr{F}}(\bm{x},t)\equiv-\bm{\partial_x\cdot}\langle\bm{uuu}\rangle-\bm{\partial_x}\langle p\bm{u}\rangle$. Applying to this the local approximation to the pressure field $p(\bm{x},t)$ that was used in the construction of the BHRZ model we have
\begin{align}
\mathscr{F}_{ij}(y,t)\approx\alpha\partial_y\langle u_i u_j u_2\rangle\equiv\alpha\partial_y R_{3,ij2},\label{Fappx}
\end{align} 
where $\alpha$ is a coefficient that determines the contributions of the self-transport and (localized) pressure-transport to the overall transport $\bm{\mathscr{F}}$. The BHRZ model for (\ref{Fappx}) is given by 
\begin{align}
\alpha\partial_y R_{3,ij2}\approx\int\limits_0^\infty \mathcal{T}^y_{ij}(y,k,t)\,dk= \frac{c_D}{2}\partial_{y}\Big(\mathscr{D}_{22}\partial_{y}{R}_{2,ij}+\mathscr{D}_{j2}\partial_{y}{R}_{2,2i}+\mathscr{D}_{i2}\partial_{y}{R}_{2,j2}\Big).\label{R3int}
\end{align} 
For the SFML with $\bm{R}_2(y,0)\propto\mathbf{I}$, $\bm{\mathcal{T}}^y(y,k,t)$ is diagonal $\forall t$, and using (\ref{R3int}) with the definition of $\bm{\mathcal{T}}^y(y,k,t)$, this leads to the result\
\begin{align}
R_{3,ij2}\approx \frac{c_D}{2\alpha}\Big(2\delta_{2i}\delta_{2j}+1\Big)\mathscr{D}_{22}\partial_y R_{2,ij},\label{R3graddiff}
\end{align} 
(index summation is not implied) where we have used the condition that in the homogeneous regions of the SFML, $\bm{R}_3=\bm{0}$ and $\partial_y \bm{R}_2=\bm{0}$. In Figure~\ref{R3_ratio} we compare results from the DNS data for $R_{3,222}/(R_{3,112}+R_{3,222}+R_{3,332})$ with the corresponding predictions following from (\ref{R3graddiff}), that is
\begin{align}
\frac{R_{3,222}}{R_{3,112}+R_{3,222}+R_{3,332}}&\approx\frac{3\partial_y R_{2,22}}{\partial_y R_{2,11}+3\partial_y R_{2,22}+\partial_y R_{2,33}},\label{R3graddiff_2}
\end{align}
where we evaluate $\partial_y\bm{R}_2(y,t)$ using the DNS data. Although the data is quite noisy, the results in Fig.~\ref{R3_ratio} seem to show that the gradient-diffusion closure gives quite a good approximation for $\bm{R}_3(y,t)$. Certainly, the relative differences between the predictions of (\ref{R3graddiff_2}) and the DNS data are insufficient to explain the much larger relative differences between the BHRZ predictions and the DNS data for the anisotropy of $\bm{R}_2(y,t)$ observed in Fig.~\ref{Anis}. This then indicates that the main source of error in the BHRZ prediction for $R_{2,22}/\mathrm{tr}[\bm{R}_2]$ arises from either the angle averaging operation applied to the evolution equation or else the local approximation for the pressure-transport\footnote[3]{There is in fact another possible cause for the discrepancies in the anisotropy results: Recall that in the DNS, the statistics are constructed by averaging over the homogeneous directions, for a given $y,t$. This leaves open the possibility that certain `phase' information remains in the statistics, that would otherwise be averaged out, or at least suppressed, under a true ensemble average (average over infinitely many flow fields for a given $y,t$), and this could affect the anisotropy results. However, we do not know how significant these phase effects are, whereas it is possible to show from the transport equations that the angle averaging operation and the local approximation for the pressure-transport will certainly affect the anisotropy predicted by BHRZ.}.

The way in which the local approximation can affect the anisotropy predicted by BHRZ can be understood as follows. In the BHRZ framework, the difference between the true and the localized form of the pressure-transport contribution to $\bm{\mathcal{N}}$, $\Delta\bm{\mathcal{P}}$, is given by  
\begin{align}
\Delta\mathcal{P}_{ij}\equiv-\nabla_i\int\limits_{\mathbb{R}^3}\Big(G(\bm{x},\bm{x}',\bm{k}) \nabla'_m\nabla'_n -\delta(\bm{x}-\bm{x}')k^{-2}k_m k_n\Big) \mathcal{R}_{3,mnj}(\bm{x}',\bm{k},t)\,d\bm{x}',\label{LPerror}
\end{align}
which represents the error introduced into the BHRZ model of the pressure-transport description because of the local approximation. Since in an anisotropic flow we would in general expect that $\mathcal{R}_{3,mn1}\neq\mathcal{R}_{3,mn2}\neq\mathcal{R}_{3,mn3}$ (and also for their gradients), then (\ref{LPerror}) shows that the error introduced to $\bm{\mathcal{R}}_2$ (and hence $\bm{R}_2$) by the local approximation will differ for different components of $\bm{\mathcal{R}}_2$. It is in this way that the local approximation to the pressure-transport will affect the ability of the model to predict the anisotropy of the fluid velocity field.

Exactly how the angle averaging and local pressure-transport approximations affect the BHRZ prediction for the generation of the flow anisotropy are, however, difficult to determine (for example we do not know the behavior of $\bm{\mathcal{R}}_{3}$ in the SFML), and a detailed investigation of these effects will be addressed in future work.
\FloatBarrier
\section{Conclusion}\label{Conc}

In this paper we have presented a comparison of the BHRZ spectral model for inhomogeneous turbulence with DNS data of a Shear-Free Mixing Layer (SFML). One of the reasons for choosing this flow is that it provides us with the opportunity to scrutinize the BHRZ model for the physical-space transport of the turbulent velocity field. We found that the model is able to capture certain features of the SFML quite well for intermediate to long-times, including the evolution of the mixing-layer width and turbulent kinetic energy. At short-times, and for more sensitive statistics such as the generation of the velocity field anisotropy, the model is less accurate. We have presented arguments that the main causes of the discrepancies are the angle averaging operations applied to the transport equations and the local approximation to the intrinsically non-local pressure-transport in physical-space that was made in the BHRZ model. Recent work \cite{rubinstein15} provides a way to asses the effects of the angle averaging operation on the model prediction of the flow field anisotropy, and this is work underway. An important next step is to incorporate non-local transport into the BHRZ model to overcome the deficiencies arising from the local approximation.

\section*{Acknowledgments}

The authors would like to thank Daniela Tordella and Michele Lovieno for kindly providing us with the data from their paper Tordella \emph{et al.}, Phys. Rev. E, vol. 77, 016309, 2008. We would also like to thank Robert Rubinstein for reading through the paper and providing helpful feedback. ADB and SK acknowledge support from the Mix and Burn project, ASC Physics and Engineering Models Program. Work at the Los Alamos National Laboratory, through the ASC Program, was performed under the auspices of the U.S. DOE Contract No. DE-AC52-06NA25396. TTC was supported by a Los Alamos National Laboratory subcontract to the University of New Mexico, No. 325696.

\appendix
\section*{Appendix}

In this appendix we discuss the discrepancies noted in Figure~\ref{Energy_func} between the BHRZ predictions and the DNS data for $y-y_c\lesssim -0.4$.
\begin{figure}[ht]
\centering
\vspace{-2mm}
\subfloat[]
{\begin{overpic}
[trim = 0mm 60mm 0mm 60mm,scale=0.4,clip,tics=20]{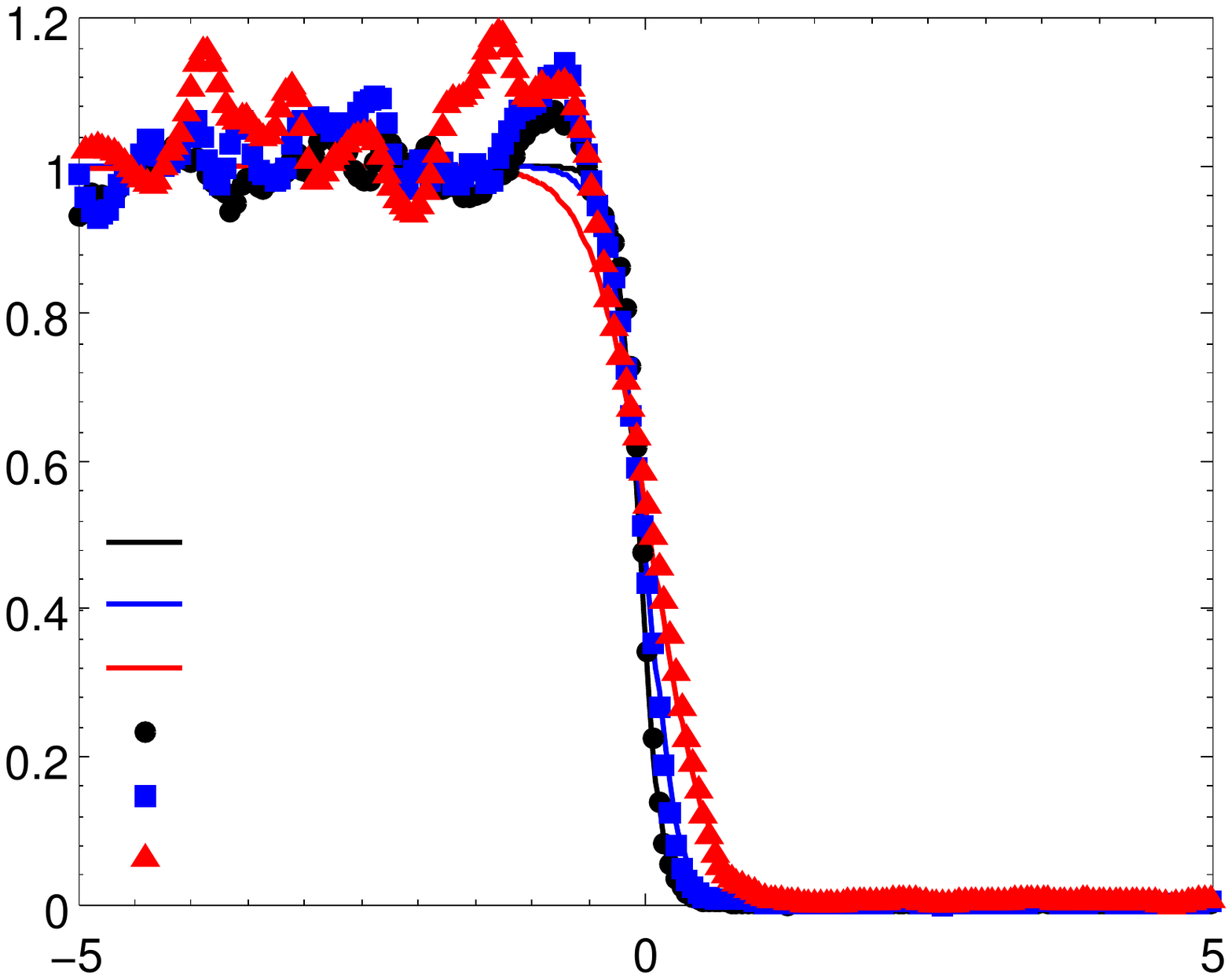}
\put(113,5){$y-y_c$}
\put(10,75){\rotatebox{90}{$\mathscr{E}(y,t)$}}
\put(55,79){\tiny{BHRZ $t/\tau=0.165$}}
\put(78,69){\tiny{$t/\tau=1.485$}}
\put(78,60){\tiny{$t/\tau=6.765$}}
\put(55,50){\tiny{DNS \hspace{2mm}$t/\tau=0.165$}}
\put(78,40){\tiny{$t/\tau=1.485$}}
\put(78,30){\tiny{$t/\tau=6.765$}}
\end{overpic}}
\subfloat[]
{\begin{overpic}
[trim = 0mm 60mm 0mm 60mm,scale=0.4,clip,tics=20]{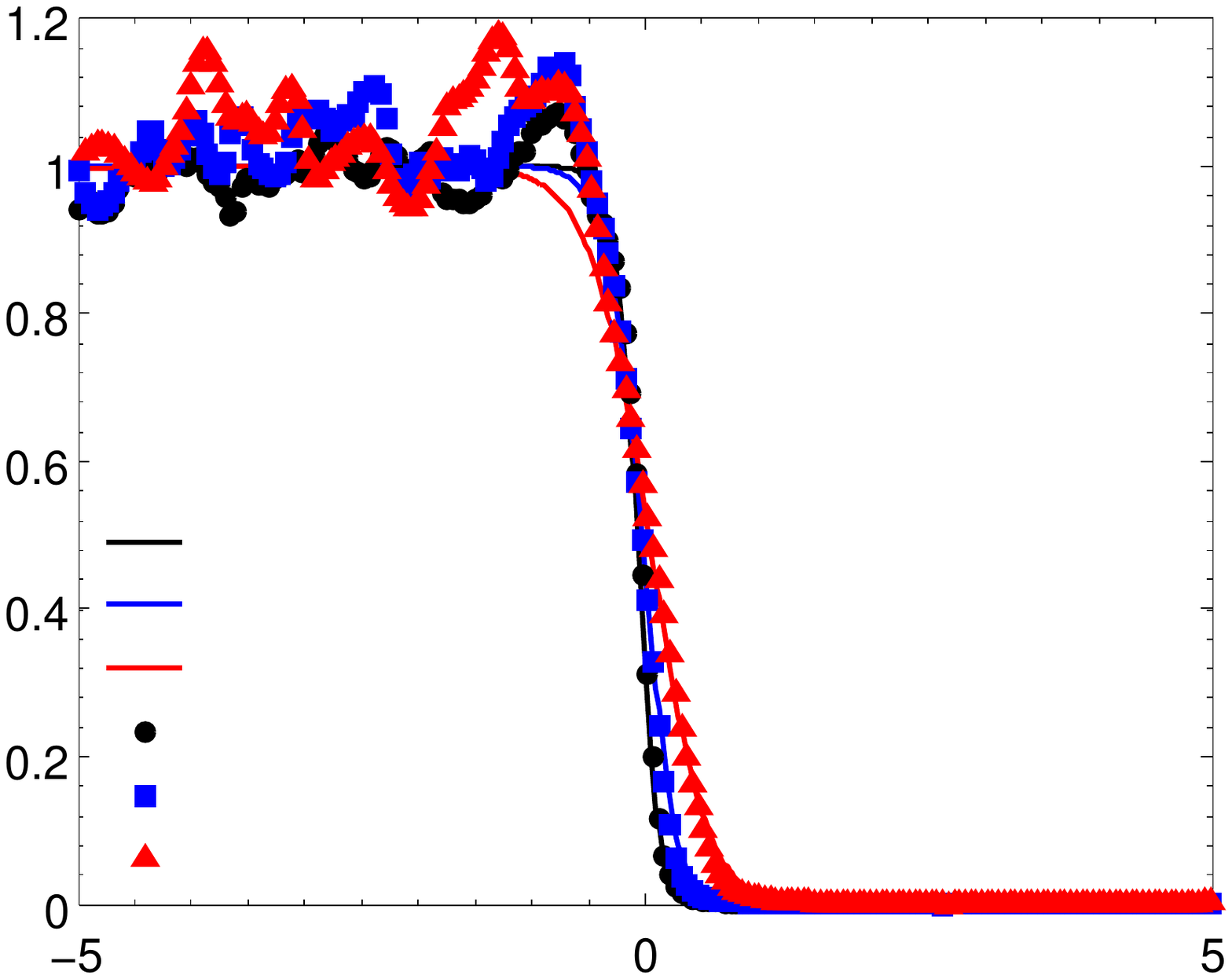}
\put(113,5){$y-y_c$}
\put(10,75){\rotatebox{90}{$\mathscr{E}(y,t)$}}
\put(55,79){\tiny{BHRZ $t/\tau=0.165$}}
\put(78,69){\tiny{$t/\tau=1.485$}}
\put(78,60){\tiny{$t/\tau=6.765$}}
\put(55,50){\tiny{DNS \hspace{2mm}$t/\tau=0.165$}}
\put(78,40){\tiny{$t/\tau=1.485$}}
\put(78,30){\tiny{$t/\tau=6.765$}}
\end{overpic}}
\caption{Plot of BHRZ predictions and DNS data for $\mathscr{E}(y,t)$ as a function of $y-y_c$ (where $y_c=L/2$) for (a) $\gamma=\sqrt{1/40}$ and (b) $\gamma=\sqrt{1/300}$. Black symbols/line correspond to $t/\tau=0.165$, red symbols/line to $t/\tau=1.485$ and blue symbols/line to $t/\tau=6.765$.}
\label{Energy_func_wide}
\end{figure}
\FloatBarrier
In plotting the DNS data for the function $\mathscr{E}(y,t)$ we must determine $\max[\mathcal{K}(y,t)]$. However, as can be seen from Figure~\ref{Energy_func_wide}, the data is noisy for $y-y_c\lesssim-0.4$ making it difficult to determine $\max[\mathcal{K}(y,t)]$. We therefore decided to choose $\max[\mathcal{K}(y,t)]$ as the value about which the data approximately oscillates at $y-y_c\leq -3$. This leads to an apparent underprediction of the BHRZ model for $\mathscr{E}(y,t)$ in the regime $-1\lesssim y-y_c\lesssim -0.4$, as seen in Fig.~\ref{Energy_func_wide}. It seems clear that this is actually a consequence of the noise in the data in this regime, rather than a genuine underprediction of the BHRZ model since the DNS results in Ireland \& Collins \cite{ireland12b}, which are also for a SFML, do not exhibit such noise and show data for $\mathcal{K}(y,t)$ (and hence $\mathscr{E}(y,t)$)  that monotonically decreases with increasing $y$ for $t\geq0$.

\bibliographystyle{unsrt}
\bibliography{refs_co12}

\end{document}